\documentclass[a4paper,11pt]{article}
\usepackage[dvips]{graphicx}
\usepackage{epsfig}
\usepackage[colorlinks=true]{hyperref}
\usepackage{pstricks}
\usepackage{amsmath,amssymb,amsfonts}
\usepackage{a4wide,epic}
\usepackage{color}

\newcommand{\demi}{\tfrac{1}{2}}
\newcommand{\q}[1]{|#1\rangle}
\newcommand{\qd}[1]{\langle #1|}
\newcommand{\as}{}

\newcommand{\highlight}[2][lightgray]{\mathchoice%
  {\colorbox{#1}{$\displaystyle#2$}}%
  {\colorbox{#1}{$\textstyle#2$}}%
  {\colorbox{#1}{$\scriptstyle#2$}}%
  {\colorbox{#1}{$\scriptscriptstyle#2$}}}%
\newcommand{\highlightt}[2][lightgray]{\colorbox{#1}{#2}}

\begin{document}
\definecolor{lightgray}{gray}{0.9}

\title{Deterministic submanifolds and analytic solution of the quantum stochastic differential master equation describing a monitored qubit}
\author{Alain Sarlette\thanks{QUANTIC lab, INRIA Paris. 2 rue Simone Iff, 75012 Paris, France; and Data Science Lab, Ghent University. Technologiepark 914, 9052 Zwijnaarde, Belgium. tel: +33 1 80 49 43 59; alain.sarlette@inria.fr}~~\&   Pierre Rouchon\thanks{Centre Automatique et  Syst\`{e}mes, Mines-ParisTech, PSL Research University.
60 Bd Saint-Michel, 75006 Paris, France. }}
\date{\today}

\maketitle

\begin{abstract}
This paper studies the stochastic differential equation (SDE) associated to a two-level quantum system (qubit) subject to Hamiltonian evolution as well as unmonitored and monitored decoherence channels. The latter imply a stochastic evolution of the quantum state (density operator), whose associated probability distribution we characterize. We first show that for two sets of typical experimental settings, corresponding either to weak quantum non demolition measurements or to weak fluorescence measurements, the three Bloch coordinates of the qubit remain confined to a deterministically evolving surface or curve inside the Bloch sphere. We explicitly solve the deterministic evolution, and we provide a closed-form expression for the probability distribution on this surface or curve. Then we relate the existence in general of such deterministically evolving submanifolds to an accessibility question of control theory, which can be answered with an explicit algebraic criterion on the SDE. This allows us to show that, for a qubit, the above two sets of weak measurements are essentially the only ones featuring deterministic surfaces or curves.
\end{abstract}



\section{Introduction}\label{sec:intro}

The quantum master stochastic differential equation is a general continuous-time model for Markovian evolution of the density operator $\rho$ (``the state'') of an open quantum system \cite{Qmeas1,Qmeas2}. Open refers to interactions with an unmonitored environment, and with measurement devices which provide continuous-time information about the quantum state. For a target system with Hilbert space $\mathcal{H}$, the effects of these interactions are characterized by operators $L_k$ on $\mathcal{H}$, sometimes called ``Lindblad operators''. Denoting
\begin{eqnarray}
\label{eq:pre-one}
F_L(\rho) & = & L \rho L^\dagger - \demi L^\dagger L \rho - \demi \rho L^\dagger L  \\ 
\nonumber
G_L(\rho) & = & L \rho + \rho L^\dagger - \mathrm{trace}(L \rho + \rho L^\dagger)\, \rho \, , 
\end{eqnarray}
the master equation reads
{\as \begin{eqnarray}\label{eq:qme}
d\rho_t & = & -i[H,\rho_t]\, dt \; + \; \sum_{k=1}^m \, F_{L_k}(\rho_t) \, dt \; + \; \sum_{k=1}^m \, G_{L_k}(\rho_t) \, \sqrt{\eta_k} \, dW^k_t \, ; \\
dy^k_t & = & \sqrt{\eta_k} \; \mathrm{trace}(L \rho + \rho L^\dagger) \, dt \; + \; dW^k_t \;\;, \;\;\; \text{for } k=1,2,...,m.
\end{eqnarray}}
This is a stochastic differential equation driven by independent Wiener processes $dW^k_t$ to be understood in the It\^o sense. The second equation describes the measurement outputs $y_k$ associated to a particular stochastic realization of the continuous weak measurement process on $m$ channels, and the first equation describes the associated stochastic evolution of the density operator $\rho$ describing the quantum state, consistent with the associated measurement output --- therefore the same noise drives both equations.
In the first equation, the first term represents Hamiltonian evolution, with $H$ hermitian. The $\eta_k$ represent measurement efficiencies. If all $\eta_k=1$, then the interaction takes place with perfectly monitored measurement devices, which lose no information about the current quantum state, and an initially pure sate ($\rho$ of rank 1) remains pure. The associated noise processes $dW^k_t$ model measurement back-action: when a quantum system is measured, its state moves closer to the stochastic result of the measurement. Completely unmonitored channels, hence often said to model interactions with ``the environment'', would correspond to $\eta_k=0$. For such channels, $\rho$ can only represent the expectation of the evolution: there is no associated measurement back-action ($G_{L_k}$ term), but still a deterministic ``relaxation'' or ``decoherence'' ($F_{L_k}$ term).\\

Our goal is to find analytic expressions for the probability distribution of $\rho_t$, for all $t>0$, evolving from a known initial state $\rho_0$, when the system follows \eqref{eq:qme}. This is to be understood as an ``\emph{a priori}'' probability distribution, i.e.~without/before knowing the sequence of measurement results actually obtained, or equivalently without knowing the values actually taken by the noise processes $dW^k_t$. The standard treatment in this case is to give the expected evolution of $\rho_t$, which is the deterministic Lindblad equation obtained with all $\eta_k=0$. This indeed describes all we can ever experimentally access if the $dW^k_t$ are unknown \emph{forever}. However, if we know that in an experimental run we will have access to the $dy^k_t$ thus the $dW^k_t$, and we can \emph{condition} some actions on these actual $dy^k_t$ as one would wish to do towards feedback control, then knowing not only the expectation but the full \emph{distribution} of future $\rho_t$ becomes relevant (e.g.~for feedback system design).

The present paper is motivated by recent experimental findings \cite{Campagne-Huard,SigZ-paper1,SigZ-paper2} on an imperfectly monitored qubit system. The qubit (quantum bit) is the smallest quantum system, with two-dimensional Hilbert space $\mathcal{H}=\mathbb{C}^2$. It has drawn a lot of attention as an illustrative benchmark of quantum control and as the basic building block for quantum information technologies \cite{NielsenChuang}. The experimental results of \cite{Campagne-Huard} have prompted us to look at \eqref{eq:qme} as a stochastic differential equation with everywhere singular diffusion. It is clear that such singular diffusion can lead to singular distributions, with lower-dimensional support --- but only for specific combinations of singular diffusion and drift. It turns out that these specific combinations hold precisely for typical quantum experiments like the one in \cite{Campagne-Huard}. Note that the lower-dimensional support manifold is in general not constant, but instead it evolves \emph{deterministically} in time. This somewhat differs from the dynamical invariants \cite{Qu-Invs}, which require the expectation of a linear operator to be invariant. Here instead, we require a nonlinear and possibly time-dependent submanifold of $\mathcal{H}$ to be determined independently of the individual realizations (with probability 1). Our first set of results, see Section \ref{sec:3.1} and Section \ref{sec:4.1}, analytically characterize the equation describing the support manifolds and their deterministic evolution in time, for typical experiments of monitored qubit systems: homodyne and heterodyne measurement with a hermitian (``quantum non demolition measurement'') or with a nilpotent (``fluorescence measurement'') operator $L_k$.
Furthermore, in Section \ref{sec:3.2} and Section \ref{sec:4.2}, we write the Fokker-Planck Partial Differential Equation (PDE) governing the evolution of the probability distribution of $\rho_t$ on this manifold, and we provide its closed-form solution. This provides a precise description of model predictions for those two particular cases, which in fact has been confronted and confirmed with experimental data in~\cite{Campagne-Huard}.

The next natural question is how general this confinement to a deterministic manifold might be for quantum systems. To answer this question, we have resorted to a control-theoretic approach which is recalled in the Appendix and might be of independent interest; indeed, we have found this method applied to a quantum system only in very reduced form in \cite{YamamotoCDC2005}. The method considers the independent noise terms as independent inputs and investigates the support of the distribution through \emph{strong accessibility} of the associated control system. The latter can be settled with a Lie algebraic criterion, i.e.~a few explicit computations determine the dimension of the support of the system at any given time, without having to actually identify the equation of the manifold.
We here develop its specific adaptation to a general quantum master equation of type \eqref{eq:qme}, and we apply it in Section \ref{sec:AllSigs} to establish that for a qubit system, the cases of Section \ref{sec:Heterodyne} and Section \ref{sec:Homodyne} are essentially the only ones where the qubit distribution remains supported on a submanifold. This points towards a small class of quantum systems which might be of particular interest for quantum control purposes.

While the present paper is devoted to a comprehensive treatment of two-level systems, we would like to mention that the algebraic characterization of submanifolds applies quite efficiently to higher-dimensional Hilbert spaces; a paper treating physically relevant examples is in preparation.


\section{The continuously monitored qubit}\label{sec:model}

The qubit has a two-dimensional Hilbert space $\mathcal{H}=\mathbb{C}^2$, which is usually identified with the span of two orthonormal vectors $\q{g},\q{e} \in \mathcal{H}$. The state of an open qubit system is then represented by the density operator $\rho$ which is a hermitian, positive semidefinite 2$\times$2 matrix of trace 1. We will use the Pauli operators $\sigma_x = \q{e}\qd{g}+\q{g}\qd{e}$, $\sigma_y = i(\q{g}\qd{e}-\q{e}\qd{g})$, $\sigma_z = \q{e}\qd{e} - \q{g}\qd{g}$. An alternative representation of the qubit state uses the so-called Bloch sphere coordinates $x,y,z$ such that $\rho = \frac{I + x\sigma_x + y \sigma_y + z \sigma_z}{2}$. The set of density matrices then corresponds to all $x,y,z$ for which $x^2+y^2+z^2 \leq 1$, also known as the Bloch sphere.

In the next two sections, we analyze in detail the evolution of the distribution of $\rho$ for a qubit model, in the four following typical experimental settings.

In \emph{homodyne detection} \emph{[Ho]}, \eqref{eq:qme} features a single monitored coherence channel $L_1$ while in \emph{heterodyne detection} \emph{[He]} it features two channels $L_1$ and $L_2 = i \, L_1$. Physically, this corresponds respectively to measuring one or the two quadratures of a field that has interacted with the qubit. Concerning the $L_1$ operator, we focus on two experimentally implemented schemes. One applies $L_1$ a hermitian operator \emph{[H]}, denoted $L_1 = \sigma_z$, which physically corresponds to a quantum non demolition  measurement; 
this is the most common type, see e.g.~\cite{SigZ-paper1,SigZ-paper2}. Another one applies $L_1$ nilpotent \emph{[N]}, denoted $L_1 = \sigma_- = \q{g}\qd{e}$ with little loss of generality, which physically corresponds to a fluorescence measurement; see e.g.~\cite{Wiseman2002} for a description of the homodyne case, with feedback, and \cite{Campagne-Huard} for an experiment involving the heterodyne case. This yields four cases to consider:
\begin{eqnarray}
\label{eq:HeH} [HeH] & & d\rho_t =  2F_{\sigma_z}(\rho_t) \, dt \; + G_{\sigma_z}(\rho_t) \, \sqrt{\eta} \, dW^1_t + G_{i\sigma_z}(\rho_t) \, \sqrt{\eta} \, dW^2_t
\\
\label{eq:HeN} [HeN]  & & d\rho_t =  2F_{\sigma_-}(\rho_t) \, dt \; + G_{\sigma_-}(\rho_t) \, \sqrt{\eta} \, dW^1_t + G_{i\sigma_-}(\rho_t) \, \sqrt{\eta} \, dW^2_t
\\
\label{eq:HoH} [HoH]  & & d\rho_t =  F_{\sigma_z}(\rho_t) \, dt \; + G_{\sigma_z}(\rho_t) \, \sqrt{\eta} \, dW^1_t
\\
\label{eq:HoN} [HoN] & & d\rho_t =  F_{\sigma_-}(\rho_t) \, dt \; + G_{\sigma_-}(\rho_t) \, \sqrt{\eta} \, dW^1_t
\end{eqnarray}

Remarkably, for each of these four typical cases, the stochastic evolution of the qubit is in fact confined to a deterministically evolving submanifold of the state space, as we show next. We start with the heterodyne cases, where the qubit stays on a deterministically moving surface. In the homodyne cases, the qubit stays on a deterministically evolving curve. This dimensional reduction in uncertainty also allows us to provide a closed-form expression for the probability distribution at any time of the qubit governed by one of \eqref{eq:HeH}-\eqref{eq:HoN}.


\section{Heterodyne measurement}\label{sec:Heterodyne}

In Bloch sphere coordinates, both dynamics \eqref{eq:HeH} and \eqref{eq:HeN} are invariant under rotation around the associated $z$ axis. Physically this just expresses that when measuring both quadratures, the reference phase for distinguishing the individual quadratures can be chosen arbitrarily. We therefore rewrite the equations in cylindrical coordinates $(r,\theta,z)$ where $x=r\cos\theta$, $y=r\sin\theta$. Note that since this is a nonlinear coordinate change it requires to carefully apply It\^o calculus\footnote{
We only need the following It\^o rule. Let $X$ a vector of components $(X)_j$ and evolving according to $(dX_t)_j = F_j(X_t) dt + \sum_k G_{j,k}(X_t)\, dW^k_t$, with normalized independent Wiener processes $dW^k_t$. Consider the change of coordinates $(X')_j = H_j(X)$. Then we have
\newline $
(dX'_t)_j = \sum_l\, \tfrac{\partial H_j}{\partial X_l} F_l dt + \sum_{l,k}\, \tfrac{\partial H_j}{\partial X_l} G_{l,k}\, dW^k + \tfrac{1}{2} \sum_{l,m,k}\, \tfrac{\partial^2 H_j}{\partial X_l \partial X_m} G_{l,k}\, G_{m,k}\, dt \, .
$\newline
The last term is the It\^o correction term, with respect to standard (i.e.~non-stochastic) calculus.}.
We get respectively:
\begin{eqnarray}
[HeH] \label{eq:HeH:Bloch} & & dr_t = -{\as (4-2\eta) \, r_t}\, dt - {\as 2 r_t} z_t \, \sqrt{\eta} dW^z_t \\
\nonumber & & dz_t = {\as 2}(1-z_t^2) \sqrt{\eta} dW^z_t\\
\nonumber & & d\theta_t = {\as 2} \sqrt{\eta} dW^\theta_t \\[2mm] \label{eq:HeN:Bloch}
[HeN]  & & dr_t = {\as\left(\frac{\eta(1+z_t)^2}{2r_t} - r_t \right)} \, dt + {\as \sqrt{\eta}} (1+z_t-r_t^2) \, dW^z_t\\
\nonumber & & dz_t = -{\as 2}(1+z_t) \, dt - {\as\sqrt{\eta}} r_t (1+z_t) \, dW^z_t \\
\nonumber & & d\theta_t = {\as\sqrt{\eta}} \frac{1+z_t}{r_t} \; dW^\theta_t \, ,
\end{eqnarray}
where $dW^r_t$ and $dW^\theta_t$ are two independent Wiener processes, equivalent to $dW^1_t$ and $dW^2_t$ via a unitary change of frame.


\subsection{Deterministic surfaces}\label{sec:3.1}

We have the following results:\vspace{2mm}

\noindent \textbf{Theorem 1:}\newline
(a) \emph{The qubit governed by the SDE \eqref{eq:HeH:Bloch} remains at all times confined to the submanifold
\begin{equation}
\highlight{[HeH]} \quad \highlight{r^2 - b_t\, (1-z^2) = 0} \quad \text{, where }\;\; \highlight{b_t = b_0 \, e^{\as -8(1-\eta) t}} \;\; \in [0,1] \, .
\end{equation}}
(b) \emph{The qubit governed by the SDE \eqref{eq:HeN:Bloch} remains at all times confined to the submanifold
\begin{equation}\label{eq:Th1b}
\highlight{[HeN]} \quad \highlight{\frac{r^2}{2} + c_t(1\text{\emph{+}}z)^2 - (1\text{\emph{+}}z) = 0} \quad \text{, where }\;\; \highlight{c_t = (c_0-\tfrac{\eta}{2})\, e^{\as 2t} + \tfrac{\eta}{2}} \;\; \in [\tfrac{1}{2},+\infty) \, .
\end{equation}}

\begin{figure}[hb]
\includegraphics[width=75mm,trim=1.5cm 1cm 1.5cm 1cm,clip=true]{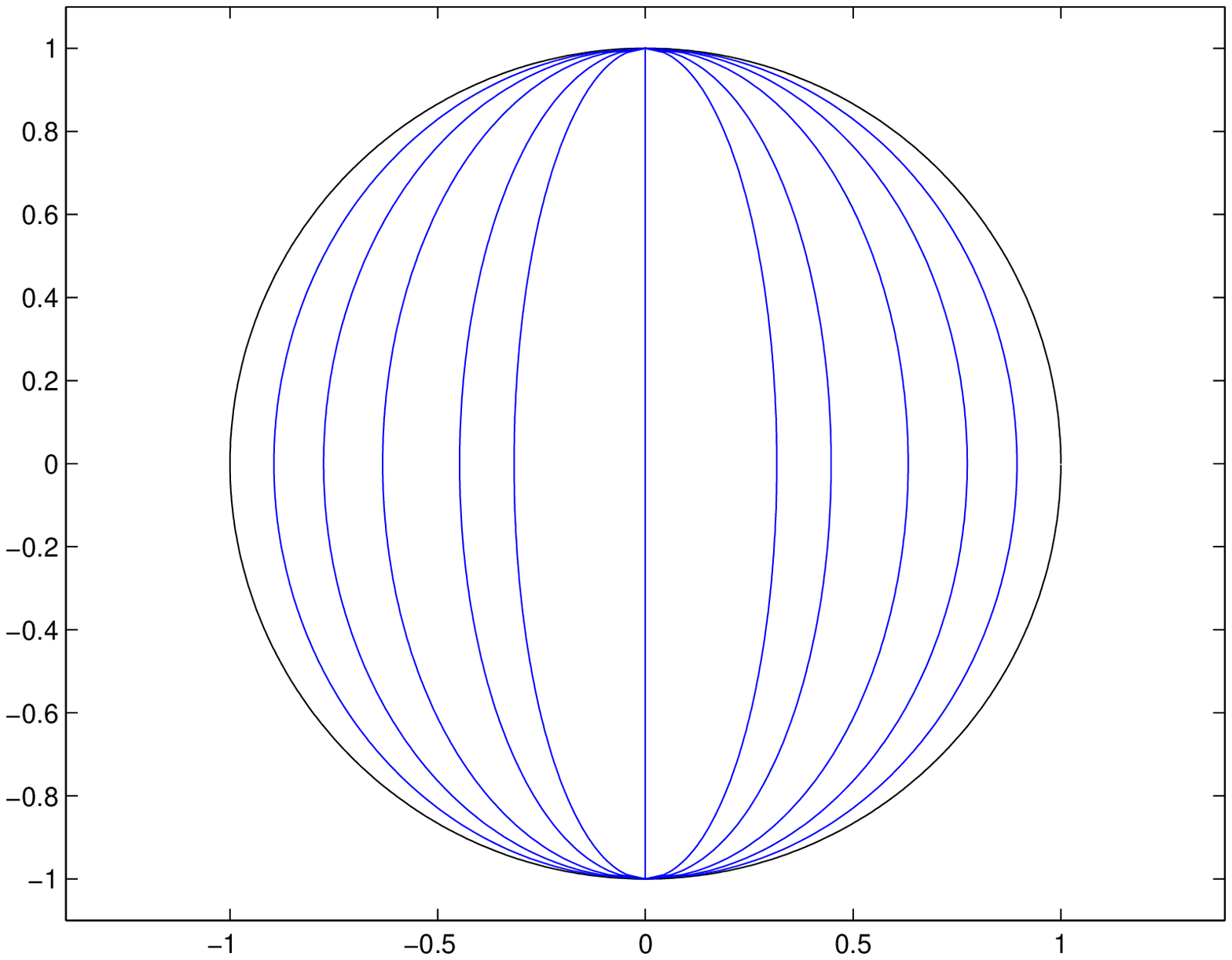}\hfill
\includegraphics[width=75mm,trim=1.5cm 1cm 1.5cm 1cm,clip=true]{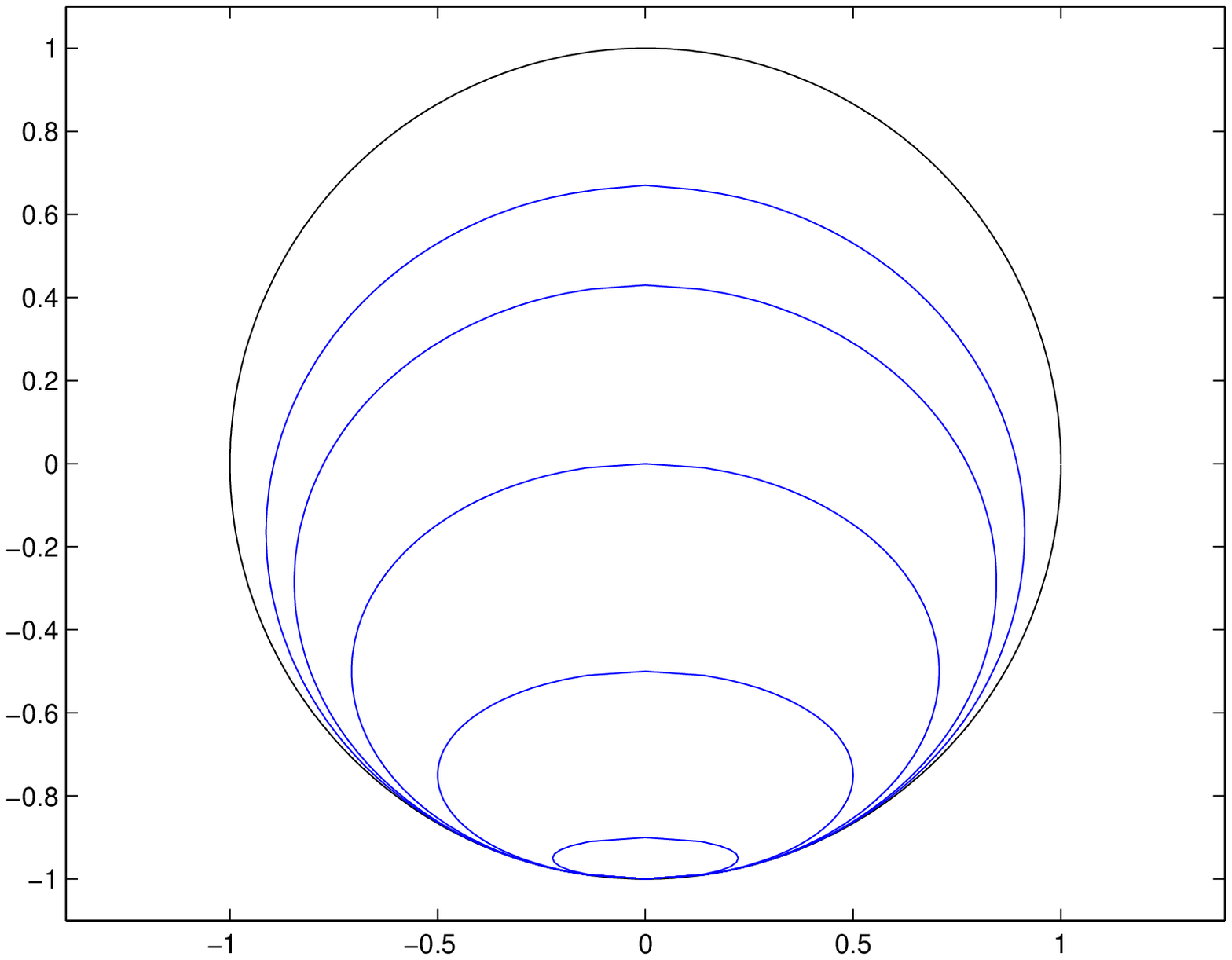}
\caption{\emph{Left:} Invariant manifolds for $\sigma_z$ measurement. I.e.~curves resulting from the intersection of the plane $x=0$ with the surface described by $r^2 - b_t\, (1-z^2) = 0$, for $b_t \in \{1, 0.8, 0.6, 0.4, 0.2, 0.1, 0 \}$ respectively from outer to inner ellipse; the outermost ellipse ($b_t=1$) is the unit circle. \emph{Right:} Invariant manifolds for $\sigma_-$ measurement. I.e.~curves resulting from the intersection of the plane $y=0$ with the surface described by $\frac{r^2}{2} + c_t(1+z)^2 - (1+z) = 0$, for $c_t \in \{0.5, 0.6, 0.7, 1, 2, 10 \}$ respectively, from outer to inner ellipse; the outermost ellipse ($c_t=0.5$) is the unit circle.}\label{fig:SurfHet}
\end{figure}

In other words, the variables $b = \frac{r^2}{1-z^2}$ and $c = \frac{1+z - r^2/2}{(1+z)^2}$ are deterministic components of the respective dynamics. The cut of the mentioned surfaces with the plane $\cos\theta=0$ are shown on Fig.~\ref{fig:SurfHet}.
Theorem 1 can be checked simply by writing the dynamics of $b_t$ and $c_t$ respectively using \eqref{eq:HeH:Bloch}, \eqref{eq:HeN:Bloch} and applying the rules of It\^o calculus. This yields a deterministic \emph{and} perfectly autonomous evolution equation, e.g.~
$\;\; dc_t = {\as 2} (c_t -\frac{\eta}{2}) \, dt \;\; \, .$
It might be more informative to explain how we have found the appropriate variables. Inspired by experimental results \cite{Campagne-Huard} for \eqref{eq:HeN:Bloch}, we have considered the surfaces defined by $\frac{r^2}{2} + f_c(z) = 0$ where $f_c$ is a function to be found, and $c \in \mathbb{R}$ parametrizes the surfaces. As a first condition, we impose that the purely stochastic part of the motion is tangent to the surface. This requires
$$r(z+1-r^2) - \frac{df_c}{dz}r(1+z) = 0 = r \left( (z+1) + 2 f_c(z) - (1+z)\frac{df_c}{dz} \right) \, .$$
The general solution of this equation is $f_c(z) = c(1+z)^2 - (1+z)$, as appearing in \eqref{eq:Th1b}. We have then checked that the associated variable $c_t$ follows an autonomous ordinary differential equation, i.e.~$dc_t$ can be expressed as a function of $c_t$ only.\\

Let us briefly discuss the interpretation of these results.
\begin{itemize}
\item For the $\sigma_-$ measurement, $c_t$ can only increase. For $c_t=1/2$ the surface coincides with the unit sphere and as $c_t$ increases the surface shrinks inwards and collapses towards $z=-1$ as $c\rightarrow +\infty$. This reflects that when monitoring a qubit via the energy loss channel $\sigma_-$, the trajectories always eventually converge towards the ground state. The only exception to this is when $c(0)-\eta/2 = 0$, in which case $c(t) = c(0)$ for all $t$. Since $c(0) \geq 1/2$ and $\eta \leq 1$, this only happens for $\eta=1$ and starting on the surface of the Bloch sphere, in which case we know from information-theoretic arguments that the quantum state remains pure. In this case, from physical arguments the state will still eventually converge to the ground state, but it will do so while remaining on the unit Bloch sphere; in this particular case, the convergence to the ground state is solely characterized by the evolution of state distribution induced by the SDE \emph{on} this invariant manifold, which is the subject of the next section. Interestingly, for $\eta=1$ but non-pure initial state i.e.~$x_0^2+y_0^2+z_0^2 < 1$, $c(t)$ further decreases. For $\eta=0.24$, the predicted evolution matches well with experimental results \cite{Campagne-Huard}. Note that the deterministic shrinking of the surface on which the trajectories lie does not preclude that the stochastic evolution drives a state from the side towards the summit of the surface, hence temporarily increasing its expectation of energy, as has indeed been observed in \cite{Campagne-Huard}.
\item For the $\sigma_z$ measurement, the surfaces are ellipsoids whose long axis coincides with the polar axis and whose short axes, of equal length $b_t$, give the radial extension; thus $b_t=1$ characterizes the unit sphere and $b_t=0$ the polar axis. As expected, $b_t$ decreases exponentially unless $\eta=1$, in which case $b_t$ remains constant. Thus unlike for $\sigma_-$, with $\sigma_z$ even if we do not start on the unit sphere, for $\eta=1$ we stay on the same surface for all times.
\end{itemize}

We conclude by mentioning the points $z=\pm 1$ for $\sigma_z$ (resp.~$z=-1$ for $\sigma_-$) which do not belong to a unique manifold $b_0$ (resp.~$c_0$). At these points the coefficients of the noise vanishes and the noiseless dynamics as well, which means that any initial state on these points will remain exactly there forever.


\subsection{Distribution on the surfaces}\label{sec:3.2}

In the following we give expressions for the probability distribution of $\rho_t$ at any time $t$, when the system starts from a known initial condition (i.e.~a Dirac distribution at some point in the Bloch sphere) and follows \eqref{eq:HeH:Bloch} or \eqref{eq:HeN:Bloch}. In the previous section we have shown that the distribution is then supported on a deterministically evolving submanifold of the Bloch sphere, and the following thus handles the remaining degrees of freedom on that submanifold.

For a more general initial state distribution, the expression of the probability distribution at any time $t>0$ is obtained simply by convolution of the results below with the distribution of initial states.\\


We recall the following link between SDE and probability distribution, for component $(X)_\ell$ of a stochastic process in $X \in \mathbb{R}^n$ evolving subject to independent Wiener processes $dW_t^k$:
\begin{eqnarray}\label{eq:rule}
& & (dX_t)_\ell = F_\ell(X_t) dt + \sum_k\, G_{\ell,k}(X_t) \; dW^k_t \;\;\;\; \text{(It\^o)}\\
\nonumber & & \Rightarrow  \;\;\; \tfrac{\partial}{\partial t} P_X(x,t) = -\sum_\ell \frac{\partial}{\partial x_\ell} (F_\ell(x) P_X(x,t)) + \tfrac{1}{2} \sum_{j,\ell} \frac{\partial^2}{\partial x_j \partial x_\ell} \left({\textstyle \sum_k  G_{j,k}(x) G_{\ell,k}(x)} P_X(x,t)\right) \, .
\end{eqnarray}
This rule allows us to write the Fokker-Planck equation in convenient coordinates, and in fact provide a closed-form solution $P_X(x,t)$ for all the cases considered in this paper. For heterodyne measurement, having confined the system to a deterministic submanifold, we have essentially to characterize a distribution with $x=(z,\theta)$.

\paragraph*{Case of $\sigma_-$:} This is in fact the most difficult case, as the evolution of $\theta$ depends on $z_t$. However nothing in the SDE depends on $\theta$ and this allows us to solve the system sequentially.

\begin{figure}
\setlength{\unitlength}{0.9mm}
\begin{picture}(85,64)
\put(0,0){\includegraphics[width=75mm,trim=1.5cm 1cm 1.5cm 1cm,clip=true]{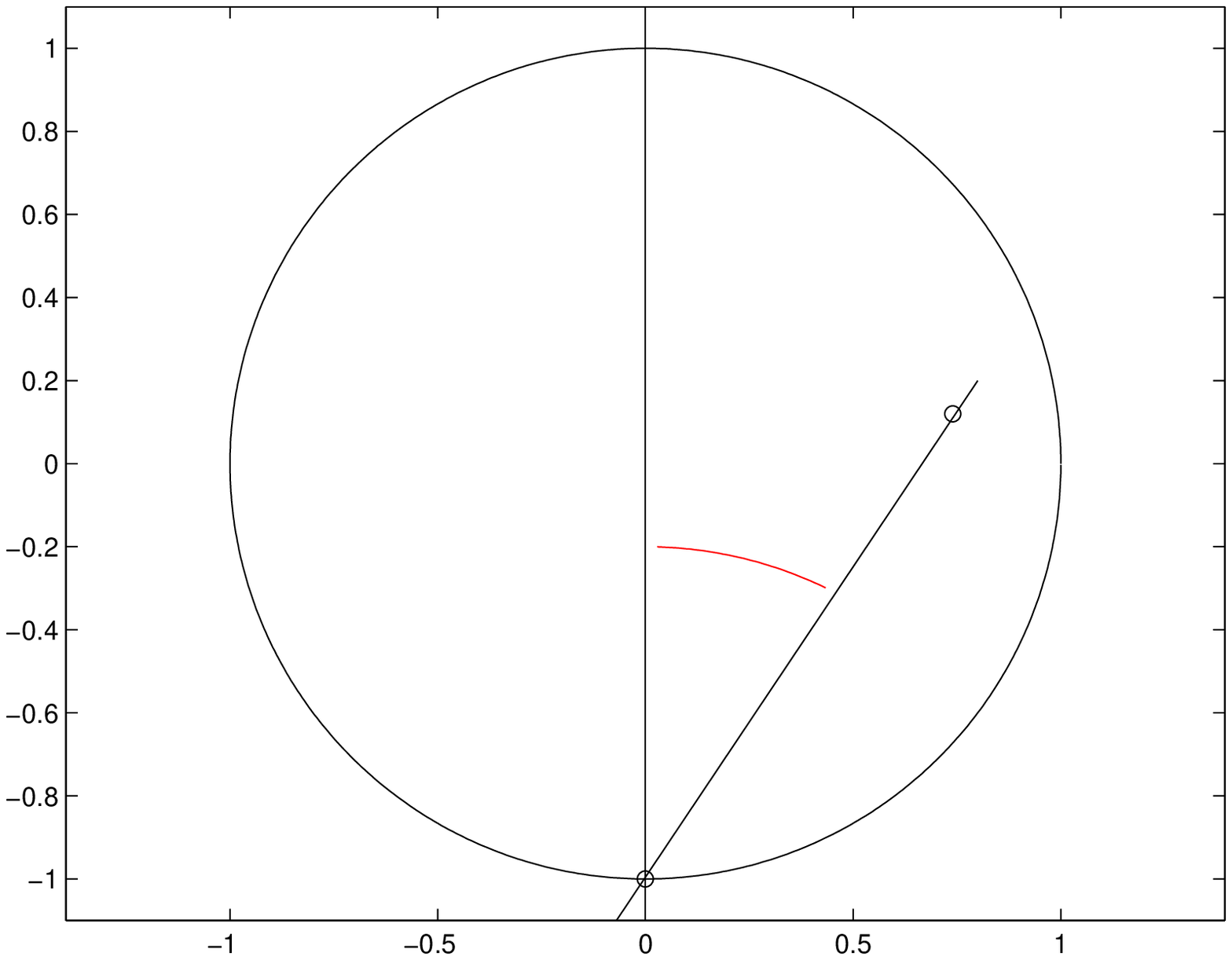}}
\put(45,4){$z=-1$}
\put(65,36){$(x,y,z)$}
\put(51,28){$\alpha$}
\put(7,59){$\phi=\tan^2(\alpha)/\eta$}
\end{picture}
\includegraphics[width=78mm,trim=1.5cm 0cm 1.5cm 0.5cm,clip=true]{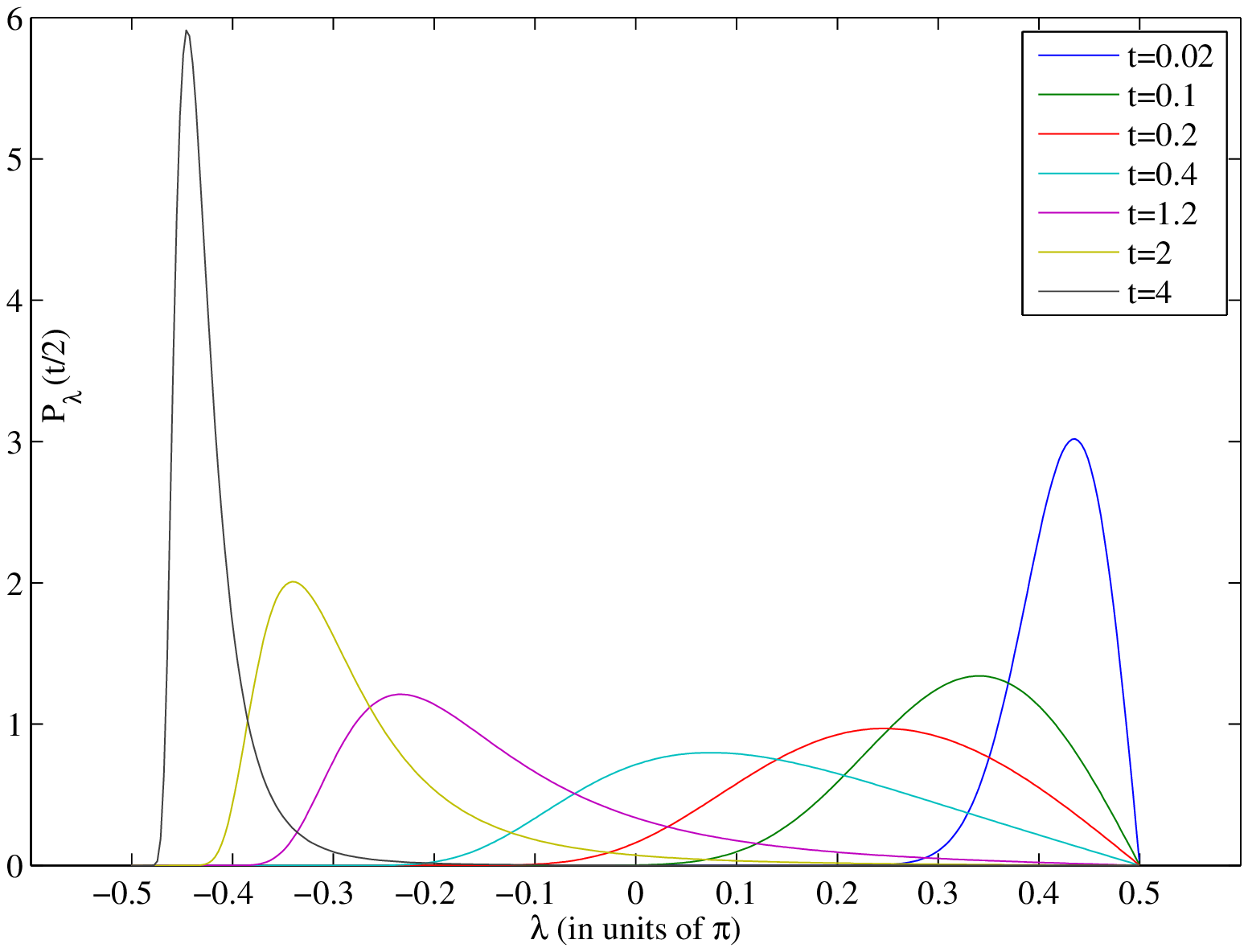}
\caption{\emph{Left:} Geometric interpretation of the variable $\phi$ introduced in \eqref{eqr:defphi}. \emph{Right:} A few snapshots of the probability distribution of type \eqref{eq:PDforphi}, i.e.~distribution of Bloch sphere latitude $\lambda$ when starting at $z=+1$ with heterodyne measurement of $\sigma_-$ and $\eta=1$.}\label{fig:lattophi}
\end{figure}

It turns out that a particularly simple expression is obtained by describing the position on the submanifolds with the variables $(\phi,\theta)$ instead of $(z,\theta)$, where
\begin{equation}\label{eqr:defphi}
\highlight{\phi = \frac{2}{\eta} \, \left(\frac{1}{z+1} - c \right) \, = \, \frac{r^2}{\eta(1+z)^2}}\, .
\end{equation}
Geometrically, $\phi$ is ($\eta$ times the square of) the tangent of the angle between the $z$ axis and the vector joining the point $z=-1$ to the state, see Figure \ref{fig:lattophi}. So $\phi$ characterizes the central projection of a state of the Bloch sphere, from the center $z=-1$ onto the plane at e.g.~$z=+1$. A state on the unit sphere close to $z=-1$ corresponds to $\phi$ close to $+\infty$, while a state close to $z=+1$ corresponds to $\phi$ close to $0$. Applying the It\^o rule and re-expressing the longitudinal diffusion, we obtain
\begin{eqnarray*}
d\phi_t & = & \sqrt{\phi_t} {\as 2}\, dW^z_t + {\as 2}\left(1+\phi_t + \frac{\eta\phi_t}{c_t+\eta\phi_t/2} \right)\, dt \, ,\\
d\theta_t & = & 1/\sqrt{\as \phi}\; dW_t^\theta  \; .
\end{eqnarray*}

We next decompose the joint probability distribution $P_X(\phi,\theta,t)$ along the Fourier modes in $\theta$:
$$
\highlight{[HeN]} \;\;\;\;\; \highlight{P_X(\phi,\theta,t) = \sum_{k=0}^{+\infty}  A_k(\phi,t) \cos(k\theta)} \; ,
$$
where we have only kept the cosines thanks to symmetry, assuming without loss of generality a solution starting at a known point with $\theta=0$. Our goal is now to get an expression for the functions $A_k(\phi,t)$. We thus use the formula \eqref{eq:rule} to get the Fokker-Planck-like equation on $P_X(\phi,\theta,t)$ and then decompose it to obtain an equation for each $A_k(\phi,t)$. The latter do not look too nice, but with the time-dependent change of variables $F_k(\phi,t) = \frac{A_k(\phi,t)}{\eta \phi + 2 c_t} \phi^{-k/2}$ we get:
\begin{equation}\label{eq:lowetaKummer}
{\as \frac{1}{2}} \frac{\partial F_k}{\partial t} = \phi \frac{\partial^2 F_k}{\partial \phi^2} + (1+k-\phi) \frac{\partial F_k}{\partial \phi} - (2+k/2) F_k \, .
\end{equation}
Towards interpretation, we note that in particular for $k=0$ we have $A_0=P_\phi$ the marginal probability distribution in $\phi$, and the $F_0$ behaves like $P_\phi$ for small $\phi$ and like $P_\phi/\phi$ for large $\phi$.

Following the standard procedure which expands the solution as a linear combination of time-exponential terms:
$F_k(\phi,{\as t/2}) = \int_s \, F_{k,s}(\phi) \, \text{exp}(-s t)\, ds\;$, the functions $F_{k,s}(\phi)$ are solutions of an ordinary \emph{differential equation of Kummer type}
\begin{equation}\label{eq:Kummer}
\phi \frac{\partial^2 F_{k,s}}{\partial \phi^2} + (b-\phi) \frac{\partial F_{k,s}}{\partial \phi} - a F_{k,s} = 0 \, ,
\end{equation}
with parameters $b=1+k$ and $a=2+k/2-s$. These Kummer type equations have a general solution in terms of regular and singular confluent hypergeometric functions. For $(2+k/2-s) = a = -n$ a negative integer, the confluent hypergeometric functions reduce to the generalized Laguerre polynomials $L^{(\alpha)}_n(\phi)$ with $\alpha=b-1=k$. The latter form a complete orthonormal basis in the sense
\begin{equation}\nonumber 
\int_0^{+\infty} \; L^{(\alpha)}_n(\phi) L^{(\alpha)}_m(\phi) \,\frac{n!}{\Gamma(n+1+\alpha)}\, \phi^\alpha\, e^{-\phi} d\phi = \left\{ \begin{array}{ll} 1 & \text{ if } n=m\\ 0 & \text{ if } n \neq m \end{array} \right. \; ,
\end{equation}
where $\Gamma$ is the gamma function that satisfies $\Gamma(n+1)=n!$ for $n\geq 0$ integer.

This allows to expand an initial condition concentrated at $\phi=\phi_0$, $\theta=0$ into
$$F_{k,s}(\phi) \propto \sum_{n=0}^{+\infty} \; a^k_n \, L_n(\phi) \;\;\; \text{with } \highlight{a^k_n = L^{(k)}_n(\phi_0)\, \phi_0^{k/2} \, \frac{n!}{\Gamma(n+1+\alpha)}} \, .$$
After computing back the change of variables, the evolution of the associated probability distribution is then characterized by:
$$\highlight{[HeN]} \;\;\;\;\; \highlight{A_k({\as t/2},\phi) = (\phi\, e^{-t})^{k/2}\; \frac{2c_0-\eta+\eta(\phi+1)e^{-t}}{N}\; \sum_{n=0}^{+\infty} \;(a^k_n e^{-(n+1)\, t}) \, L^{(k)}_n(\phi)} \, ,$$
where $N$ is a normalization constant. The series is well-behaved as soon as $t > 0$, since for fixed $\phi$ the $L^{(k)}_n(\phi)$ (and thus their product with the $a_n$) are bounded uniformly in $n$. For large $t$, the term with $n=0$ becomes dominant, hence we converge towards a uniform distribution in $\phi$, which implies that in terms of latitude on the Bloch sphere, the state concentrates more and more towards the point $z=-1$ (see Figure \ref{fig:lattophi}). Also $k=0$ becomes dominant for large $t$, reflecting that the $\theta$ variable gets uniformly distributed as well.

In some cases the above series can be worked out more explicitly. For instance we can compute the marginal distribution $P_Z(z,t)$ when starting at $z=+1$, which means $\phi_0=0$ and $c_0=1/2$. We then have $a^0_n = 1$ for all $n$ and using the property $\sum_{n=0}^{+\infty} d^{n+1} L^{(0)}_n(x) = \tfrac{d}{1-d} \, e^{\frac{-d}{1-d} x}$ we get
$
P_\phi(\phi,{\as t/2}) = (1-\eta+\eta(\phi+1)e^{-t}) \, d_t\; e^{-d_t \phi} \;\;\; \text{with} \;\;\; d_t = \frac{1}{e^t-1} \, .
$
For $\eta=1$, in terms of latitude $\lambda$ on the Bloch sphere where $z=\sin(\lambda)$, this corresponds to the distribution represented on the right of Fig.\ref{fig:lattophi}:
\begin{equation}\label{eq:PDforphi}
P_\lambda({\as t/2},\lambda) = \frac{4 \cos(\lambda)}{(1+\sin(\lambda))^3}\, e^{-t}\, d_t\; e^{-d_t\,(1-\sin(\lambda))/(1+\sin(\lambda))} \;\;\; \text{with} \;\;\; d_t = \frac{1}{e^t-1} \, .
\end{equation}

\paragraph*{Case of $\sigma_z$:} For the hermitian measurement operator, the dynamics on the surface is described by two decoupled purely diffusive evolutions, for the $z$ and $\theta$ coordinates:
\begin{eqnarray*}
d\theta_t & = & {\as 2}\sqrt{\eta} dW^\theta_t \\
dz_t & = & {\as 2} \sqrt{\eta} \, (1-z_t^2)\, dW^z_t \, .
\end{eqnarray*}
Thus for an experiment starting at any known initial state on the Bloch sphere, at any time the associated probability distribution is the product of the marginal distribution $P_Z$ in $z$ and the marginal distribution $P_\theta$ in $\theta$.

Regarding $\theta$, we have a canonical Brownian motion on the circle. The solution can be expanded in the Fourier basis as
\begin{eqnarray*}
P_\theta(t,\theta) & = & \sum_{k=-\infty}^{+\infty} \; a_k(t) \, e^{i k\theta}
\quad \text{with } a_k(t) = e^{-{\as 2}k^2\eta t}\, a_k(0) \, .
\end{eqnarray*}
In principle any initial distribution $P_\theta(0,\theta)$ can be expanded on the Fourier basis functions; the Dirac distribution, corresponding to a known starting position, gives $a_k(0) = 1/(2\pi)$ for all $k$. As soon as $t>0$ the series converges fine and for large $t$, $P_\theta(t,\theta)$ converges towards the uniform distribution. Alternatively, one can write the Gaussian solution for a Brownian motion on the line and wrap it around the circle to get the fast converging series:
$$\highlight{[HeH]} \quad \highlight{P_\theta({\as t/4},\theta) = \frac{1}{\sqrt{2\pi \eta t}} \sum_{k \in \mathbb{Z}} \, e^{-(\theta - \theta_0 + 2\pi k)^2/ 2 \eta t}} \, .$$
\vspace{2mm}

\begin{figure}
\includegraphics[width=78mm,trim=1.5cm 0cm 1.5cm 0.5cm,clip=true]{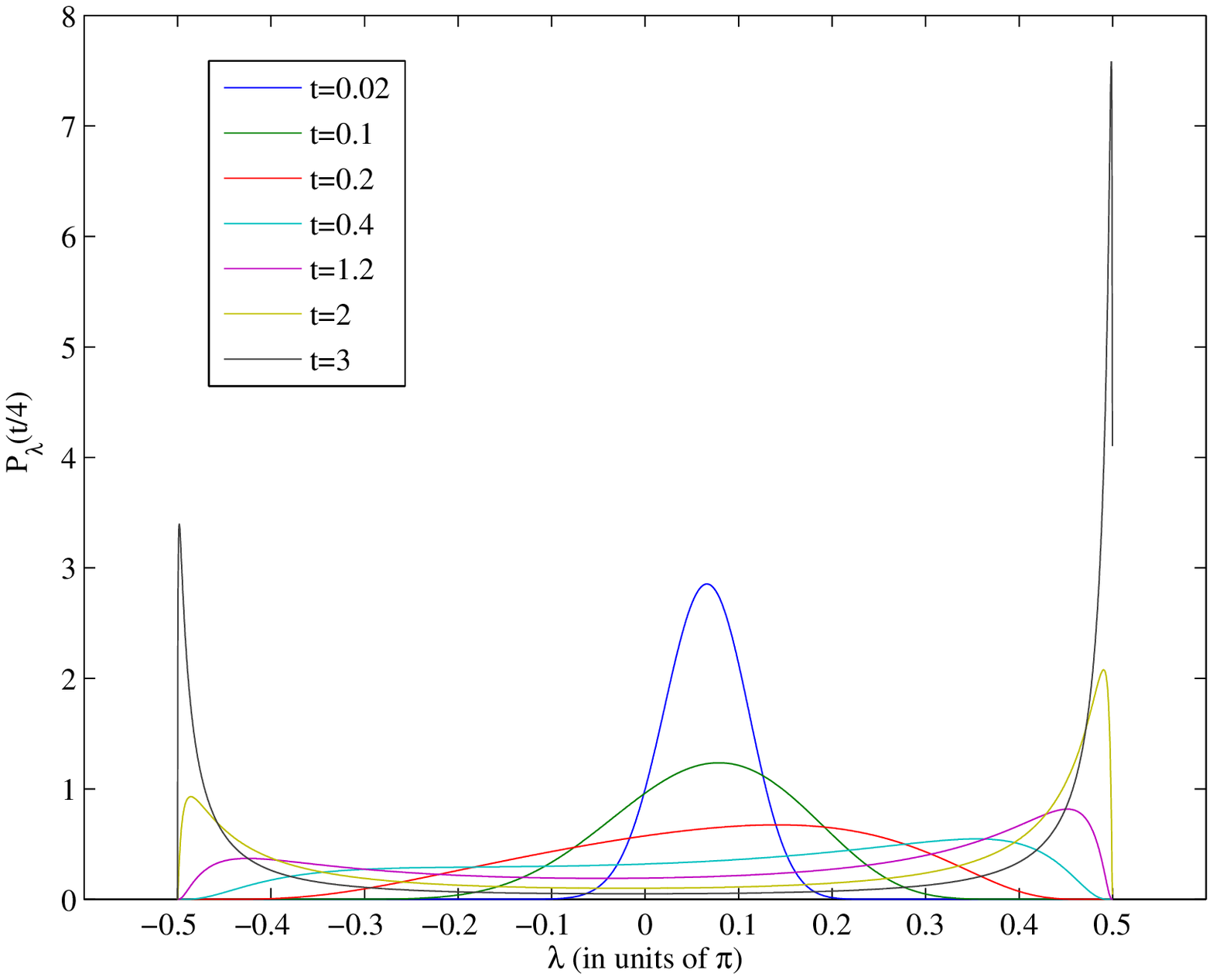}
\includegraphics[width=78mm,trim=1.5cm 0cm 1.5cm 0.5cm,clip=true]{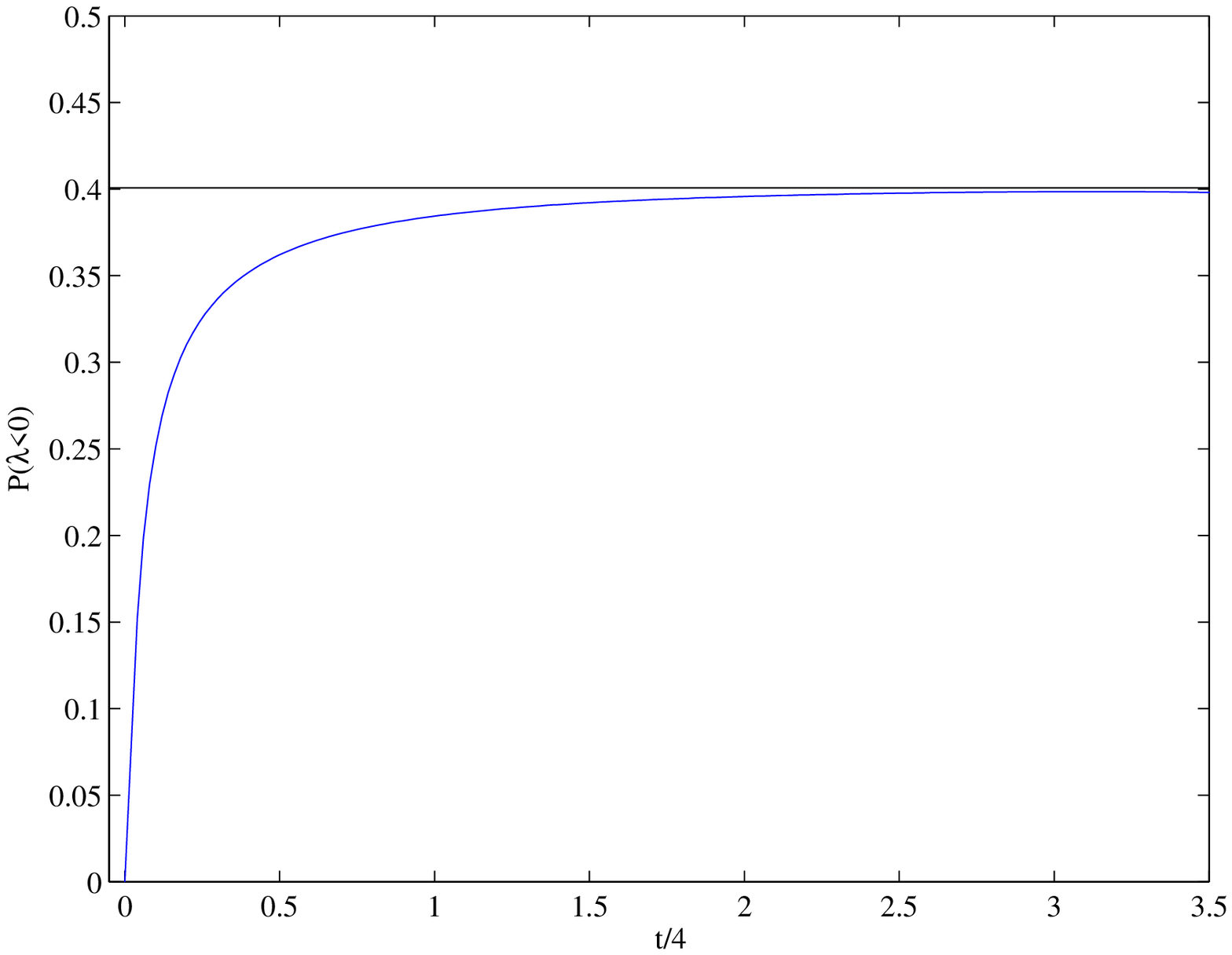}
\caption{\emph{Left:} A few probability distributions of type \eqref{eq:HehPw} for the heterodyne measurement of $\sigma_z$, converted into probability distributions on the Bloch sphere latitude $\lambda = \text{asin}(z)$ for $\eta=1$. The system starts slightly above the Bloch sphere equator, at $\lambda_0=0.2$. \emph{Right:} Evolution in time of the weight of probability in the southern hemisphere of the Bloch sphere, for the same data as the left plot. The horizontal line depicts the fraction of detections $z=-1$ expected in a projective measurement of $\sigma_z$.}\label{fig:3}
\end{figure}

To solve the diffusive SDE in $z$, we use the change of variables $\highlight{w=\text{asinh}(\frac{z}{\sqrt{1-z^2}})}$, such that $w$ is close to the tangent of the latitude on the Bloch sphere $\tan\lambda = \frac{z}{\sqrt{1-z^2}}$. This maps the interval $z \in [-1,1]$ onto $w \in [-\infty,+\infty]$ and we have the Fokker-Planck equation:
$$\frac{\partial P_w}{\partial t} = {\as 2 \eta} \frac{\partial^2 P_w}{\partial w^2} - {\as 4} \eta \frac{\partial}{\partial w}\left(\text{tanh}(w) \, P_w \right) \, .$$
Defining $P_w({\as t/4},w) = \text{cosh}(w) Q_w(t,w) e^{-\eta\,t/2}$ we get
$$\frac{\partial Q_w}{\partial t} = \frac{\eta}{2} \frac{\partial^2 Q_w}{\partial w^2} \, .$$
This is the Fokker-Planck equation of a canonical Brownian motion, whose Gaussian solution starting at $w=w_0$ writes
$$Q_w(t,w) =  \frac{1}{\sqrt{2\pi \eta t}} \, e^{-(w - w_0)^2/ 2 \eta t} \, .$$
Converting back to $P_w$,
the whole expression simplifies considerably and we are just left with two Gaussians with opposite drifts:
\begin{equation}\label{eq:HehPw}
\highlight{[HeH]} \quad\highlight{P_w({\as t/4},w) = \tfrac{1}{2\text{cosh}(w_0)\, \sqrt{2\pi \eta t}} \, \left( e^{w_0}\, e^{\frac{-(w-w_0-\eta t)^2}{2\eta t}} + e^{-w_0}\, e^{\frac{-(w-w_0+\eta t)^2}{2\eta t}} \right) \, .}
\end{equation}
This expresses exactly how the distribution converges towards the steady states $z=+1$ and $z=-1$, which would be the possible results for a \emph{projective} measurement of $\sigma_z$. The two Gaussians indeed separate, since their drift is in $t$ while their standard deviation grows in $\sqrt{t}$. Moreover, one checks that $\;\; \frac{e^{w_0}}{e^{-w_0}} = \frac{1+z_0}{1-z_0} \;\; ,$ i.e.~the relative weight between the Gaussians matches the relative probabilities of detecting $+1$ or $-1$ in a projective measurement of $\sigma_z$. See figure \ref{fig:3} for illustration.

\section{Homodyne measurement}\label{sec:Homodyne}  

For the $\sigma_z$ measurement, the symmetry around the $z$ axis of the Bloch sphere remains. Translating \eqref{eq:HoH} into cylindrical coordinates we get
\begin{eqnarray}
[HoH] \label{eq:HoH:Bloch} & & dr = - {\as 2}r \, dt - {\as 2}r z \, \sqrt{\eta}\, dW_z \\
\nonumber & & dz = {\as 2}(1-z^2) \sqrt{\eta}\, dW_z\\
\nonumber & & d\theta = 0 \; .
\end{eqnarray}
For the homodyne $\sigma_-$ measurement, the symmetry is broken since the noise part is reduced to a $\sigma_- + (\sigma_-)^\dagger = \sigma_x$ operator, without the same on $i\,\sigma_- + (i\sigma_-)^\dagger = \sigma_y$. 
Hence it is easier to keep the Cartesian Bloch sphere coordinates,
\begin{eqnarray}
[HoN] \label{eq:HoN:Bloch} & & dx = -x/2\, dt + \sqrt{\eta}((1+z) - x^2)\, dW_t \\
\nonumber & & dy = -y/2\, dt + \sqrt{\eta} (-xy) \, dW_t \\
\nonumber & & dz = -(1+z)\, dt + \sqrt{\eta} (-x(1+z)) \, dW_t \, .
\end{eqnarray}


\subsection{Deterministic curves}\label{sec:4.1}

We have the following deterministically evolving components.\vspace{2mm}

\noindent \textbf{Theorem 2:}\newline
(a) \emph{The qubit governed by the SDE \eqref{eq:HoH:Bloch} remains at all times confined to the curve defined by
\begin{equation}\label{eq:ellip}
\highlight{[HoH]} \;\; \highlight{\theta_t = \theta_0}\;\; \text{ and }\;\; \highlight{r^2 - b_t\, (1-z^2) = 0} \;\;, \text{ where }\;\; \highlight{b_t = b_0 \, e^{\as -4(1-\eta) t}} \;\; \in [0,1] \, .
\end{equation}}
(b) \emph{The qubit governed by the SDE \eqref{eq:HoN:Bloch} remains at all times confined to the curve defined by the intersection of
\begin{eqnarray}
\highlight{[HoN]}\;\;\; \highlight{\frac{x^2}{2} + c_t(1\text{\emph{+}}z)^2 - (1\text{\emph{+}}z) = 0}  &  \text{ where }& \highlight{c_t = (c_0-\frac{\eta}{2})\, e^{t} + \frac{\eta}{2}} \;\; \in [1/2,+\infty) \; \\
\text{ and }\;\; \highlight{y - f_t(1\text{\emph{+}}z) = 0} & \text{ where }& \highlight{f_t = f_0 \, e^{t/2}}\;\; \in [0,+\infty) \; .
\end{eqnarray}}

Like for the heterodyne measurement, this can be checked just by writing the dynamics of the proposed coordinates and checking that (i) they are influenced by no noise components directly and (ii) they obey an \emph{autonomous} differential equation. The presence of a single noise component is a necessary condition for having the distribution of states supported, at any time, on a \emph{curve} in the Bloch sphere; however as we will see in Section \ref{sec:AllSigs} this is not sufficient in general, due to condition (ii). The existence of deterministic curves is thus due to our particular choice of $L_1$ operators.

Regarding the geometry of the support curves:
\begin{itemize}
\item For $\sigma_z$ we have a straightforward variation of the heterodyne case, i.e.~the support is the curves represented on Fig.~\ref{fig:SurfHet}(left) and contained in the plane $\theta = \theta_0$. {\as Note that in the heterodyne case, $\sigma_z$ and $i\sigma_z$ were both giving the same drift, and as we are now left with only one of those channels, the evolution of $b_t$ -- i.e.~the contraction orthogonal to the direction associated to the measurement output -- is  two times slower. In contrast, the $z$ variable is precisely the measurement output associated to $\sigma_z$, while $i\sigma_z$ was anyways giving pure noise as measurement output: therefore it is not surprising that the evolution of $z_t$ in \eqref{eq:HoH:Bloch} keeps the same speed as in the heterodyne case.}
\item For $\sigma_-$, the equation associated to $c_t$ describes the same shape as for the heterodyne case, associated to the curves represented on Fig.~\ref{fig:SurfHet} in the plane $x=0$. However, while for the heterodyne case the surface is generated by rotation of such curve around the $z$ axis, for the present homodyne case it is obtained by translating the curve along the $x$ axis. At each time, the deterministic curve is the intersection of this surface with the surface associated to $f_t$, and which corresponds to a plane parallel to the $x$ axis and through the point $z=-1$. As $f_t$ increases in time, those planes move from vertical towards horizontal. Note that the $f_t$ evolution is \emph{independent of $\eta$} and features two static surfaces: a stable one at $f_t=+\infty$, whose intersection with the Bloch sphere is the single state $z=-1$; and an unstable one at $f_t=0$, corresponding to the plane $y=0$ which includes both $z=-1$ and $z=+1$. The intersection of the time-dependent surfaces associated to $c_t$ and to $f_t$ gives a curve with $z$ maximum in the plane $x=0$, and getting more and more horizontal.

This is consistent with an information-theoretic interpretation: only variations in the $x$ direction, through the associated information gain, might imply that energy increases over some parts of the trajectory, while transversely to the $x$ direction (variable $f_t$) the energy decreases deterministically.
\end{itemize}
The points $z=\pm 1$ for $\sigma_z$ (resp.~$z=-1$ for $\sigma_-$) are still the only points which do not correspond to unique values of $b_0$ (resp.$c_0, f_0$); and they are still equilibrium points, where the coefficient of the noise vanishes and the state will stay forever.


\subsection{Distribution on the curves}\label{sec:4.2}

\paragraph*{Case of $\sigma_z$:} The remaining stochastic differential equation is like for the heterodyne case,
$$dz_t = {\as 2} \sqrt{\eta} \, (1-z_t^2)\, dW^z_t \, .$$
The same change of coordinates and solution are thus valid: with $\;\highlight{w=\text{asinh}(\frac{z}{\sqrt{1-z^2}})}\;$,
$$\highlight{[HoH]} \;\;\; \highlight{P_w({\as t/4},w) = \tfrac{1}{2\text{cosh}(w_0)\, \sqrt{2\pi \eta t}} \, \left( e^{w_0}\, e^{\frac{-(w-w_0-\eta t)^2}{2\eta t}} + e^{-w_0}\, e^{\frac{-(w-w_0+\eta t)^2}{2\eta t}} \right)} \; .$$
\vspace{2mm}

\paragraph*{Case of $\sigma_-$:} Unlike the manifolds, the probability distribution is similar to the heterodyne setting, just without the variable $\theta$. Since the marginal distribution on $z$ now contains all the information, in fact we follow the steps of the corresponding heterodyne case without the $\theta$-dependent expansion, taking directly $k=0$.

We start from the stochastic differential equation
$$dz_t = -(1\text{+}z_t)\, dt - \sqrt{2\eta}\,\sqrt{(1\text{+}z_t) - c_t (1\text{+}z_t)^2}\, (1\text{+}z_t)\, dW_t$$
which is obtained by replacing $x_t$ according to Theorem 2(b). This is the similar replacement as for $r_t$ in the heterodyne case, so we try a similar change of variables:
$$\highlight{\textstyle \chi = \frac{x^2}{2\eta\, (1+z)^2} = \left(\frac{1}{1+z} - c_t \right) \frac{1}{\eta}} \, .$$
This yields
$$d\chi_t = \sqrt{2}\sqrt{\chi_t}\, dW_t + \left( \chi_t + \frac{1}{2} + \frac{2\eta\chi_t}{c_t + \eta \chi_t} \right) \, dt$$
and defining again $F_\chi = P_\chi / (c_t + \eta \chi)$ we get
$$\frac{\partial F_\chi}{\partial t} = \chi \frac{\partial^2 F_{\chi}}{\partial\chi^2} + (3/2-\chi) \frac{\partial F_{\chi}}{\partial\chi} - 2 F_{\chi} \, .$$
Pursuing the same approach again leads to a Kummer type differential equation \eqref{eq:Kummer} with now $a=2-s$ and $b=3/2$. For $a=-n$ a negative integer, the solutions of this equation include the generalized Laguerre polynomials $L_n^{(\alpha)}(\chi)$ with $\alpha=1/2$.

Thus an initial condition concentrated at $\chi=\chi_0$ can be expanded as
$$F_s(\chi) \propto \sum_{n=0}^{+\infty} \; a_n \, L^{(\alpha)}_n(\chi) \;\;\; \text{with } \highlight{a_n = L^{(\alpha)}_n(\chi_0) \, \frac{n!}{\Gamma(n+3/2)}}$$
and the associated probability evolves as
$$\highlight{[HoN]}\;\;\; \highlight{P_\chi(t,\chi) = \tfrac{c_0-\eta/2+\eta(\chi+1/2)e^{-t}}{N}\; \sum_{n=0}^{+\infty} \;(a_n e^{-(n+1)\, t}) \, L^{(\alpha)}_n(\chi)} \, ,$$
with $N$ a normalization constant. 


\section{When is a qubit confined to a deterministic surface or curve?}\label{sec:AllSigs}

The fact that the SDEs \eqref{eq:HeH} to \eqref{eq:HoN} feature deterministically evolving submanifolds has appeared as a surprise to us. It is certainly not implied directly by the fact that they are governed by less than 3 noise channels. Indeed, even if there is a single noise term, its combination with deterministic drift can spread out the state of an SDE into several dimensions. This is essentially the same principle of nonholonomic motion that allows a car to park sidewards: by concatenating different steps associated to different values of a single degree of freedom in the vector field (the steering angle), one can reach all positions and orientations in the plane. Thus for a generically selected SDE, one would rather expect that associated nonholonomic constraints cannot be integrated, and diffusion indeed takes place in all directions.

Given the observations on \eqref{eq:HeH} to \eqref{eq:HoN}, a natural question is whether the confinement to submanifolds is a particularity of the chosen Lindblad operators, or maybe a more general property of quantum dynamics. It turns out that it is the former, as we next show for the qubit. Towards better readability, we only summarize the results in this section. Their proof is based on tools from control theory, leading to an algebraic criterion to analyze directly from a vector field in how many dimensions a system can spread. This theory is recalled in the Appendix, which also discusses its specific adaptation to quantum systems.

This general theory allows in just a few systematic minutes to retrieve the \emph{existence} of the deterministic submanifolds of Sections \ref{sec:Heterodyne} and \ref{sec:Homodyne}, without constructing their explicit form. For general choices of measurement operators $L_k$, we will first consider the case when all the channels are monitored. We then illustrate a few relevant cases with unmonitored channels. Our general assumptions are:
\begin{itemize}
\item[(i)] We can start from any point inside the Bloch sphere, i.e.~we do not consider cases implied by special initial conditions (like a pure state).
\item[(ii)] The relative strengths of the different Lindblad operators are not precisely tuned, i.e.~we do not consider cases where the drift implied by one operator would precisely cancel an effect of another operator.
\item[(iii)] Similarly, the values of the $\eta_j$ are not precisely tuned towards cross-cancelations, except for the case $\eta=1$ which we sometimes treat independently.
\end{itemize}

\subsection{One-dimensional submanifolds}\label{sssec:1D}

\noindent \textbf{Proposition 3:} \emph{Consider a qubit subject to a single arbitrary Lindblad operator $L$. The state remains confined to a deterministic time-varying curve in the Bloch sphere if and only if
$$\;\; [L,L^{\dagger}]\, L\;\; = r L + c I \;\;\quad \text{ for some $r \in \mathbb{R}$ and $c \in \mathbb{C}$.}$$
 This condition holds iff
$\;\; \highlight{L= c_1\, \sigma_-}\;\;$ or $\;\; \highlight{L = c_1 \sigma_z + c_2 I} \;\;$ in some orthonormal Hilbert basis and for some $c_1,c_2 \in \mathbb{C}$.}
\vspace{2mm}

\noindent \underline{Proof:} According to the theory in Appendix, staying on a curve imposes on the Lie algebras generated by the vector fields, that $\mathfrak{G}_F = \mathfrak{G} = \text{span}(G_{L_1})$. Obviously, in presence of any unique arbitrary operator $L$, we have $\mathfrak{G}=\text{span}(G_{L})$ since $[L,\,L] = 0$. Incorporating the drift term towards $\mathfrak{G}_F$ imposes that the vector field generated by $[L,L^{\dagger}]\, L$ must be collinear with the vector field generated by $L$.

A general operator $L$ can be written as $L=\alpha \sigma_z + \beta i \sigma_x + \gamma i \sigma_z + c I$ for $\alpha,\beta,\gamma \in \mathbb{R}$ and $c \in \mathbb{C}$, without loss of generality modulo a unitary change of basis. We then get
\begin{eqnarray*}
[L^{\dagger},L]\, L & = & [2\alpha \sigma_z,L]\, L
= -4 \alpha \beta \, \sigma_y\, L \, .
\end{eqnarray*}
Writing it out, we see that having this last expression collinear with $L$ requires either $\beta=0$, or $\alpha=0$, or $L=\alpha(\sigma_z \pm i\sigma_x)$. The latter case is essentially $L=c_1\, \sigma_-$; indeed recall that $\sigma_- \propto (\sigma_x + i \sigma_y)$ and that $e^{i\phi} \sigma_- = U\, \sigma_- U^\dagger$ with $U=\text{diag(}e^{i\phi},\,1)$ a unitary change of basis. The other cases are, up to basis change, of the form $L = c_1 \sigma_z + c_2 I$.
\hfill $\square$\\

The cases of Section \ref{sec:Homodyne} are readily checked.$\;\;$ For $L=\sigma_-$ we have $[\sigma_+,\sigma_-]\, \sigma_- = -\sigma_-$. $\;\;$ For $L=\sigma_z$ we get  $[\sigma_z,\sigma_z]\, \sigma_z = 0$. The latter is recovered as the second case in Proposition 4, with $c_2=0$ and $c_1$ real. For $c_1$ imaginary we get a Hamiltonian rotation, at stochastic speed, around the $z$ axis. For general complex $c_1$, the system moves on the surface described by \eqref{eq:ellip} and undergoes a stochastic rotation proportional to its latitude change, hence remaining on an inclined curve inscribed in that surface.\vspace{2mm}

\noindent \textbf{Ex.1.a:} Interestingly, the above shows that e.g.~for $L=\sigma_z + \sigma_-$, the system would \emph{not} remain confined to a curve, even though we have a single noise channel. One checks indeed that
\begin{eqnarray}\label{eq:Ex1now}
[L^{\dagger},\,L]\, L & = & -L - 2 i \sigma_y \,  =: -L + 2 L' \, .
\end{eqnarray}

\subsubsection{Several Lindblad operators}

With the restrictions imposed by Prop.3, it is not difficult to see that staying on a deterministic time-varying curve in presence of several Lindblad operators requires very special conditions.\vspace{2mm}

\noindent \textbf{Proposition 4:} \emph{Consider a qubit subject to Lindblad operators $L_j$, $j=1,2,...$. The state remains confined to a deterministic time-varying curve in the Bloch sphere if and only if
$$\highlight{L_j = c_0 \sigma_z + c_j I} \;\; \text{for all $j$ in some basis and for some $c_0,c_j \in \mathbb{C}$.}$$}

\noindent \underline{Proof:} We again use the algebraic criteria from the appendix. To keep the Lie algebra $\mathfrak{G}$ one-dimensional, we already need $L_j = L_1 + c_j\, I$ with some $c_j \in \mathbb{C}$. Combining this with the still necessary requirements for a single $L_1$, we see that the case $L_1=\sigma_x+i\sigma_y$ drops out (unless $L_j=L_1$ for all $j$ which is not really ``several'' operators). For the only remaining case, the properties \eqref{eq:GQrhoA} and $G_{L+\alpha I} = G_L$ ensure that indeed, the system would remain confined to a deterministic curve.
\hfill $\square$

\subsection{Two-dimensional submanifolds}

Let us first clarify when a single monitored channel $L$ can confine the system to a two-dimensional submanifold of the Bloch sphere.\vspace{2mm}

\noindent \textbf{Proposition 5:} \emph{Consider a quantum master SDE \eqref{eq:qme} on a qubit with $H=0$ and a single, monitored operator $L$.
\begin{itemize}
\item If $\highlight{\eta<1}$, then the system starting at a generic $\rho(0)$ will diffuse in all 3 dimensions of the Bloch sphere, except for the cases in \highlightt{Prop.3}, where it stays on a deterministic curve.
\item If $\highlight{\eta=1}$ and either $\highlight{L = \sigma_x + i \beta \sigma_y + r\, I}$ in some basis, with any $\beta,r \in \mathbb{R}$, or \linebreak $\highlight{L = \sigma_x + i \sqrt{1+r^2} \sigma_y + r \, i I}$ in some basis, with any $r \in \mathbb{R}$, then the system starting at any $\rho(0)$ remains confined to a 2-dimensional deterministically evolving submanifold. Else, even for $\eta=1$, except for the cases in Prop.3 the system starting at a generic $\rho(0)$ diffuses in all 3 dimensions of the Bloch sphere.
\end{itemize}}
\vspace{2mm}

\noindent \underline{Proof:} Again we write a generic $L$ and examine the algebraic conditions in the appendix. Throughout the proof we exclude the cases already covered in Section \ref{sssec:1D}. Note that for $\eta<1$, as we assume that $\eta$ is not precisely tuned, both the part in $\eta$ and the part in $(1-\eta)$ of the drift \eqref{eq:GQrho} must be considered independently towards building $\mathfrak{G}_F$. In particular the vector fields associated to the following commutators belong to $\mathfrak{G}_F$ as constructed with Prop.B.6:
\begin{eqnarray*}
L & = & \alpha \sigma_x + \beta i \sigma_y + \gamma i \sigma x + c I\\ 
\, [L^\dagger,\, L]\, L =: L' & \propto & \beta \sigma_x - \gamma \sigma_y + c \sigma_z + \alpha i \sigma_y\\
\, [L,\,L'] & \propto & -\beta c \sigma_x + \gamma c \sigma_y + (\beta^2 + \gamma^2 - \alpha^2) \sigma_z - \alpha c i \sigma_y - 2 \alpha \gamma i \sigma_z \, .
\end{eqnarray*}
Their linear dependence is examined with Proposition B.5. Note that $c$ can be complex, so terms must be slightly regrouped. Satisfying the first option of Prop.B.5 requires $\mathrm{Im}(c)=0$, then $\gamma=0$ and $(\beta^2-\alpha^2+|c|^2)=0$. The second option of Prop.B.5 gives no new solution. The third option is the most general and allows to have just $\gamma=\mathrm{Im}(c)=0$ or $\gamma=\mathrm{Re}(c)=0$. It then remains to check the Lie bracket between $G_{L'}$ and $F_L+D_L$.
\newline - For $\eta=1$, this reduces to investigating the commutator $Q_2 = \left[L'\,,\, (L^\dagger+L)\, L \right]$. For $c$ real we get
$$ Q_2 \in \text{span}(\sigma_x,\,\sigma_z,\,i \sigma_y) \, .$$
Such operator obviously satisfies the third item of Prop.2.A together with $L$ and $L'$. Hence $G_{Q_2}$ can be expressed as a linear combination of $G_{L'}$ and $G_{L}$, so the whole $\mathfrak{G}_F = \mathrm{span}(G_{L'},G_{L})$. Thus: $L = \sigma_x + i \beta \sigma_y + r I$ is good.
\newline - For $\eta=1$ and $c=i\,r$ imaginary we get
$$ Q_2 \propto \alpha \beta \sigma_x + (\beta^2-r^2) i \sigma_y + \alpha r\, i \sigma_z\, .$$
In this case we have $G_{Q_2} \in \text{span}(G_{L},G_{L'})$ only when $\beta^2 = r^2 + \alpha^2$.
\newline - For $\eta<1$, we have additional terms in $[F_L+D_L,\,G_{L'}]$. For $c$ real, the $G_{Q_1}$ term in \eqref{eq:GQrho} admits the same property as for $G_{Q_2}$, but the remaining term belongs to $\text{span}(G_{L},G_{L'})$ only when $L$ takes one of the forms leading to a 1-dimensional manifold (see Prop.3). This can be checked rather easily by writing out the dynamics on the Bloch coordinates and checking the determinant of the 3$\times$3 matrix concatenating (i) this vector field, (ii) $G_L$ and (iii) $G_{L'}$.
\newline - For $\eta<1$ and  $c=i\,r$ imaginary, the $Q_1$ is as well parallel to $L'$ so the $G_{Q_1}$ term of \eqref{eq:GQrho} lies in the already generated subspace of $\mathfrak{G}_F$. And again, the remaining term with $[L,L']\rho L^\dagger$ adds a third direction to the Lie algebra, unless $r=0$.
\hfill $\square$\\

\noindent \textbf{Ex.1.b:} Let us pursue Ex.1.a and check whether for $L=\sigma_z + \sigma_-$, the system might be confined to a 2-dimensional time-dependent submanifold of the Bloch sphere for $\eta=1$ (for $\eta<1$ we already know from Prop.5 that this is not possible). The answer is positive, since $L = (\sigma_z + \sigma_x/2) + i \sigma_y/2$ would fit one of the cases of Prop.5 after a rescaling by $\sqrt{5/4}$ and a rotation around the $y$ axis that brings $\sqrt{4/5}(\sigma_z + \sigma_x/2)$ onto $\sigma_x$. We can also double-check concretely that this operator satisfies the algebraic rank constraints. Resuming from \eqref{eq:Ex1now}, we get

$$\left[L,\,L'\right] =  L + \tfrac{1}{2} L' - \tfrac{5}{2} \sigma_x \, .$$
We thus already have $\mathrm{span}(G_{\sigma_x}, G_{i\sigma_y}, G_{\sigma_z}) \subseteq \mathfrak{G}_F$. This matches the form of the third case in Prop.B.5 for a two-dimensional vector field, i.e.~if $\mathfrak{G}_F$ contains no further element then the system would evolve on a 2-dimensional surface.

For $\eta=1$, this is confirmed by noting that $(L^\dagger+L)L = \alpha \sigma_x + \beta \sigma_z + r I$ for some $\alpha,\beta,r \in \mathbb{R}$. Thus examining $[F_{L}+D_{L},\,G_{L_k}]$, which by \eqref{eq:GQrho} is essentially determined by $2Q_2=[L_k,(L^\dagger+L)L]$, with $L_k \in \{ \sigma_x, i\sigma_y, \sigma_z \}$ yields no new terms in $\mathfrak{G}_F$. The equation of the corresponding submanifold is given by \eqref{eq:StrSurf}.

For $\eta<1$, the same applies to $2Q_1=[L_k,L^\dagger L]$. The remaining term is more cumbersome to treat; a numerical test, at a random $\rho$, readily shows that the generated vector field adds a third direction; so the system diffuses in all directions.
$\square$\vspace{2mm}

\noindent \textbf{Ex.2:} According to Prop.5, the single operator $L = \sigma_- + i \, \sigma_x$ would spread the state in 3 dimensions, even for $\eta=1$. One easily generates a few low-order Lie brackets from $\mathfrak{G}_F$ and checks, on a random $\rho$, that they indeed yield 3 linearly independent motion directions. $\square$

\subsubsection{Several Lindblad operators}

Let us now move to dynamics involving two independent operators $L_1$ and $L_2$. Examining this with the tools of the Appendix involves the algebra generated by the commutator between $L_1$ and $L_2$. An easily treated example is the heterodyne measurements of Section \ref{sec:Heterodyne}.\vspace{2mm}

\noindent \textbf{Ex.3}: Consider a heterodyne measurement, with two measurement operators $L$ and $i\, L$. Since $[L,\,iL]=0$ and $F_L = F_{i\, L}$, the possible motions are just the vector sum of those generated in the homodyne case with $L$ and with $i\, L$. Thus when homodyne measurement of $L$ keeps the system on a deterministic curve (case of Prop.3), the corresponding heterodyne measurement keeps the system on a 2-dimensional submanifold. For $\eta=1$, other homodyne measurements might confine the system to a 2-dimensional submanifold (Prop.5); but one checks that for those case, the associated heterodyne measurement in fact always spreads the state in 3 dimensions.
$\square$\vspace{2mm}

Finally, we can investigate which other pairs of measurement operators $L_1,L_2$ might keep the qubit system on a deterministic submanifold. It turns out that, at least for $\eta<1$, only the two heterodyne measurement cases of Section \ref{sec:Heterodyne} remain.
\vspace{2mm}

\noindent \textbf{Proposition 6:} \emph{Consider a quantum master SDE \eqref{eq:qme} on a qubit with $H=0$, several monitored operators $L_k$ and $\highlight{\eta_k \in (0,1)}$, starting at an arbitrary initial state $\rho(0)$. The qubit state $\rho(t)$ at each time $t>0$ will be restricted to a deterministically evolving 2-dimensional manifold if and only if one of the following conditions is satisfied in some orthonormal Hilbert basis:
\begin{itemize}
\item There exist $\beta_k,\alpha_k \in \mathbb{C}$ such that
$\;\; \highlight{L_k = \beta_k \, \sigma_z + \alpha_k I} \quad \text{for all } k.$
\item There exist $\beta_k \in \mathbb{C}$ such that
$\;\; \highlight{L_k = \beta_k \, \sigma_-} \quad \text{for all } k.$
\end{itemize}}

\noindent \underline{Proof:} The drift is now the sum of two terms $F_{L_1} + F_{L_2}$. Recall that we assume the relative strength of $L_1$ and $L_2$ is not precisely controlled. Therefore we can consider each drift independently, i.e.~in the algebraic conditions of the appendix, $\mathfrak{G}_F$ is the smallest Lie algebra containing $\mathfrak{G}$ and closed under Lie brackets with both $F_{L_1}$ and $F_{L_2}$ (instead of just with their sum). In particular, $\mathfrak{G}_F(L_1)\subseteq \mathfrak{G}_F$ and $\mathfrak{G}_F(L_2)\subseteq \mathfrak{G}_F$, so with Prop.5 for $\eta<1$ we are already restricted to
$$U_j L_j U_j^\dagger \in \{ c_1 \sigma_z + c_2 I,\, \sigma_- \vert c_1,c_2 \in \mathbb{C} \} $$
with some unitary changes of coordinates $U_j \in SU(2)$. Investigating those cases (this is a bit tedious, see Appendix C), we in fact find that no other pair $L_1,L_2$ than the ones listed above satisfies the Lie algebra condition.
\hfill $\square$\\

\noindent \textbf{Ex.4:} When $\highlight{\eta_k=1}$ for all $k$, the commutation formula based on $G_{Q_2}$, together with Prop.B.5, allows to identify further possibilities as listed in Appendix C, Proposition C.8, for which the state would remain on a 2-dimensional submanifold. This holds for instance for $L_1=i\sigma_1$ and $L_2=i\sigma_2$, with any two zero-trace Hermitian operators $\sigma_1$ and $\sigma_2$ (case [C]). One easily checks, via It\^o calculus, that the associated deterministic surface is simply
$$d(x^2+y^2+z^2) \; = \; 0$$
which is not really surprising: the motion implied by such measurement operators is just perfectly observed stochastic Hamiltonian rotation. Obviously in this case, having any number of measurement operators $L_k = i \sigma_k$, with the $\sigma_k$ zero-trace Hermitian, will not change this property. However for $\eta<1$, due to non-commuting actions if $[\sigma_1,\sigma_2] \neq 0$, the state would diffuse in all three dimensions. Physically, this expresses that as the Hamiltonian motion is not perfectly observed, the uncertainty on rotation angle makes the qubit (``information'') state converge towards the rotation axes. $\square$

\subsection{Adding unmonitored channels}

For completeness we can briefly mention an analysis with unmonitored channels, $\eta_j=0$. We consider the physically typical phase decoherence channel $L_3=\sigma_z$ in addition to the measured $L_1=\sigma_-$ (and $L_2=i\sigma_-$, in the heterodyne case).\vspace{2mm}

\noindent \textbf{Proposition 7:} \emph{Consider a quantum master SDE \eqref{eq:qme} on a qubit with $H=0$ and an unmonitored operator $L_1 = \sigma_z$ with $\eta_1 = 0$.
\begin{itemize}
\item For a heterodyne measurement with $L_2=\sigma_-$ and $L_3 = i\sigma_-$, $\eta_2,\eta_3 \in (0,1)$, the state will diffuse in all 3 dimensions of the Bloch sphere.
\item For a homodyne measurement with $L_2=\sigma_-$ and $\eta_2 \in (0,1)$, the qubit state at each time $t>0$ will be restricted to a deterministically evolving 2-dimensional manifold.
\end{itemize}}

\noindent \underline{Proof:} We again resort to the algebraic criteria from the appendix. Unlike in the previous section, the Lie algebra $\mathfrak{G}$ is generated only by $\sigma_-$ (and $i\sigma_-$, in the heterodyne case), not involving $\sigma_z$. Towards computing $\mathfrak{G}_F$, we must first apply the general formula \eqref{eq:GQrho} of Prop.B.6, with $k\in \{2,3\}$ and $j \in \{1,2,3\}$. For $j=2,3$ we have checked in Prop.3 that the Lie brackets vanish. For $j=1$, note that there is no It\^o vs.~Stratonovitch correction since there is no associated noise ($\eta_1=0$). Furthermore the term in $Q_2$ vanishes since $\eta_1=0$; the term in $Q_1$ vanishes as $[L_k,L_1^\dagger L_1] = [L_k,\,I]=0$. The remaining term, for $j=1$ and $k=2$, yields a vector field
$$[F_{\sigma_z},\,G_{L_2}] := \tilde G :\; (dx,\,dy,\,dz) = (z-1-x^2,\,-xy,\,x-xz)\, .$$
The latter is not collinear with $G_{L_2}$.\newline
- For heterodyne measurement, at a generic point of the Bloch sphere, this spans a three-dimensional space together with $G_{\sigma_-}$ and $G_{i\sigma_-}$.\newline
- For homodyne measurement, we must check further Lie brackets. Since $\tilde G$ is not of the form $G_{L}$ for some $L$, we cannot apply \eqref{eq:Gcommute} nor Prop.B.5 directly. Hence we just look directly at the vector fields in Bloch coordinates. We get $[\tilde{G},\,G_{L_1}] \in \text{span}(\tilde{G},G_{L_1})$ and also $[F_{\sigma_z},\,\tilde{G}] \in \text{span}(\tilde{G},G_{L_1})$.\hfill $\square$\\

\section{Conclusion}

In this paper we give analytic expressions for the distribution of state ($\rho$) of a qubit system governed by specific quantum Stochastic Differnetial Equations. The success of our analysis relies on the fact that the system remains confined to a lower-dimensional, deterministically evolving manifold for (almost) all realizations of the noise processes. Hence in a second step, we identify and adapt to the case of open quantum systems, a system-theoretic tool which allows to systematically check whether such confinement of the system occurs. We apply this method to the qubit case and conclude that essentially, the two main situations explored in recent experiments are the only ones where such deterministic confinement occurs. The system-theoretic method, recalled in appendix, is richer than the ones that characterize invariants of quantum systems, because we can identify when the system remains on a submanifold even if (and in some sense, independently of the fact that) this manifold is time-varying. For instance for the qubit under $\sigma_-$ measurement, the manifold deterministically collapses towards the ground state.

We must stress that our characterization of Lindblad operators for which the system remains on a deterministic submanifold assumes a generic starting point $\rho(0)$, generic values of the $\eta_k$ and of the relative strengths of the different operators. When one restricts the system for instance to start on the surface $x^2+y^2+z^2=1$ of the Bloch sphere, these conditions are \emph{not} necessary, in particular those proposed in Proposition B.5. Nevertheless, the algebraic approach proposed in appendix remains fully applicable and adaptable --- we just want to avoid singling out too many sub-cases. 

Future work should investigate how confinement to submanifolds is featured in higher-dimensional Hilbert spaces. Preliminary computations show that, at least for some experimentally relevant cases, such confinement still appears and allows to extract useful insight e.g.~towards model reduction and/or parameter identification.

The \emph{a priori} description of quantum evolutions proposed here is more informative than the traditional solution $\rho$ of the deterministic Lindblad equation \cite{Lindblad}, which just represents the average state of the system over the noise realizations. Our results show that the distribution around this average state can take quite particular forms. Such characterization can be of practical importance towards more efficiently estimating how the quantum system behaves in all situations where it is partially observed. In particular, it should help analyze and design systems where decisions will be taken conditionally on such observations, which include estimation of model parameters \cite{Six-et-al} and of course feedback stabilization. More generally, it might indeed be important towards error correction or information protection to know that the state, although not confined to a protected subspace, remains on a sort of protected nonlinear submanifold.

\section*{Acknowledgments}

The authors thank Mazyar Mirrahimi, Benjamin Huard, Philippe Campagne-Ibarcq, Pierre Six and Landry Bretheau for many useful discussions about the quantum physical aspects and the behavior of the qubit. They also wish to thank Jean Lévine for pointing out the references that have efficiently led us towards the Stroock-Varadhan theorem. This work was partially supported by the Projet Blanc ANR-2011-BS01-017-01 EMAQS (Agence Nationale de la Recherche).


\section*{Appendix: algebraic analysis of the vector fields associated to quantum SDEs}

Although everywhere-singular diffusion appears special in the context of noise-driven processes, in a control engineering context the motion achievable with a reduced number of inputs has been well characterized. In the present section we want to review how the powerful tools developed for control engineering allow to readily check whether a quantum stochastic differential equation (SDE) contains hidden deterministically evolving submanifolds.


\subsection*{A. The classical case}

The main idea is to consider the noises as --- incidentally random --- control inputs. This can be formally justified by the Stroock-Varadhan theorem \cite{Stroock-Varadhan}, which we recall here as adapted from \cite{Chaleyat-Maurel}.
\vspace{2mm}

\noindent \textbf{Proposition A.0:} \emph{Consider a stochastic differential equation
\begin{equation}\label{eq:FGst}
dx_t =  F(x_t) \, dt + \sum_{j=1}^m \, G_j(x)\, \circ\, dW^j_t \, ,
\end{equation}
with $x_t \in \mathbb{R}^N$ the state, $dW^1_t,dW^2_t,..., dW^m_t$ independent Wiener processes, $x_0$ fixed and the dynamics to be understood in the Stratonovitch sense (we therefore use the $\circ$ symbol). The support of the distribution of $x_t$ can be described as the closure, for the natural Banach topology on $C([0,1],\mathbb{R}^N)$, of the set of solutions of the following controlled system:
\begin{equation}\label{eq:FG}
d\tilde{x}_t =  F(\tilde{x}_t) \, dt + \sum_{j=1}^m \, G_j(\tilde{x})\, du^j_t \, ,
\end{equation}
with $\tilde{x}_0=x_0$, for all possible control signals $u^1_t,u^2_t,..., u^m_t$ in $H^1([0,1],\mathbb{R}^m)$. \newline (The important point is that the control signals can take values in an open set around $0$.)}
\vspace{2mm}

This link is central because the support of a controlled system \eqref{eq:FG}, especially its dimension, has been well characterized in the literature. We are interested precisely in cases where this support is of dimension $N-n$ for some $n>0$, indicating the presence of $n$ deterministic coordinates.\\

We first need to recall a few definitions.
These are rather standard and can be found in textbooks like~\cite{KalCrit}.
Given two vector fields $F$ and $G$ on the same manifold, their Lie bracket $[F,G]$ corresponds to the Lie derivative of $F$ along $G$ minus the Lie derivative of $G$ along $F$. The result is a vector field which needs not be parallel to neither $F$ nor $G$, and it gives an idea of motion directions generated by alternating infinitesimal motions with $F$ and $G$.
\vspace{2mm}

\noindent \textbf{Definition A.1:} \emph{The \emph{Lie algebra} $\mathfrak{G}$ generated by vector fields $G_1,G_2,...,G_m$ is the smallest algebra, closed under Lie brackets, that contains $G_1,G_2,...,G_m$. In other words, it is the vector space spanned by the $G_1,G_2,...,G_m$ and by all vector fields that are obtained by iterating Lie brackets of these elements. Finally, the \emph{drift-preserved Lie algebra} $\mathfrak{G}_F$ is the smallest Lie algebra containing $\mathfrak{G}$ and closed under Lie brackets with $F$, i.e.~for any $G \in \mathfrak{G}_F$ we have $[F,G] \in \mathfrak{G}_F$. Note that $F$ itself need not belong to $\mathfrak{G}_F$.}
\vspace{2mm}

Consider the controlled system \eqref{eq:FG}. From a given initial condition $x_0$, by selecting different control signals $u^1_t,u^2_t,..., u^m_t$, the user can steer the system towards different states $x_t$ at time $t$. It is clear that if the $G_j(x)$ span $\mathbb{R}^N$ at almost every $x$ (thus $m\geq N$), then the user can steer the state in any direction. However, even for $m<N$, the system can still reach any state if the Lie algebra $\mathfrak{G}$ generated by $G_1,G_2,...,G_m$ spans $\mathbb{R}^N$. This indeed indicates that by smartly alternating between $G_1,G_2,...,G_m$, new directions of motion can effectively be obtained. This is the principle allowing a car to effectively move sidewards by using the well-known parking maneuver.
The drift $F$ alone does not imply control capabilities. However, combining this drift with control fields might further help to steer the system in an unactuated direction; e.g.~an airplane whose translation velocity is fixed can also move ``sidewards'' with respect to its reference trajectory.

When the control inputs represent noise processes, deterministic evolution of some coordinates correspond to cases where the noise \emph{cannot} induce effective motion in all directions. More precisely, the concept that we examine is \emph{strong accessibility}.\vspace{2mm}

\noindent \textbf{Definition A.2:} \emph{The system \eqref{eq:FG} is \emph{strongly accessible} at a point $x_0$ if the set of all its possible solutions $x_t$ at time $t>0$, over all possible choices of the signals $u^j$, forms an open subset of $\mathbb{R}^N$. It is \emph{strongly accessible in $N-n$ dimensions} if the set of possible solutions all lie in, and form an open subset of, an $N-n$ dimensional submanifold of $\mathbb{R}^N$.}
\vspace{2mm}

In this definition the set of all possible $x_t$ does not have to be a neighborhood of $x_0$: whilst spreading out along all dimensions, the state might also be subject to a drift. Replacing controls by noises according to Proposition A.0, the $N-n$ dimensions are those in which the state \emph{effectively} diffuses stochastically (e.g.~dimension of the set of points that the car with Wiener as a driver can reach). The possible drift allowed by the definition expresses that the manifold supporting the state distribution need not be invariant in time, but might evolve deterministically.

The following result from control theory allows to assess the number of strongly accessible dimensions of a controlled system from algebraic properties of its vector fields~\cite{KalCrit}.\vspace{2mm}

\noindent \textbf{Proposition A.3:} \emph{The system \eqref{eq:FG} with analytic functions $F,G_1,...,G_m$ is strongly accessible at $x_0$ if and only if the Lie algebra $\mathfrak{G}_F$ associated to these vector fields has full dimension $N$ at $x_0$. Moreover, if $\mathfrak{G}_F$ has dimension at most $N-n < N$ for all $x_0$, then the system stays on a (time-dependent) manifold of dimension $N-n$, independently of the control inputs.}\vspace{2mm}

Proposition A.0 translates this directly to stochastic differential equations driven by Wiener processes in Stratonovitch form.


\subsection*{B. The quantum case}

In principle, Proposition A.3 can be applied directly to check whether a given quantum stochastic differential equation on a finite-dimensional Hilbert space features hidden deterministically evolving coordinates. The vector fields $F,G_j$ exactly match the expressions in \eqref{eq:pre-one}, only with $x \in \mathbb{R}^N$ replaced by $\rho \in \mathbb{R}^{N \times N}$ which is still a vector space.

A few specificities of the quantum case can be addressed to facilitate its use. In the following, spans or linear combinations are meant with \emph{real} coefficients whenever unspecified.

\subsubsection*{B.1 Quantum operators vs.~vector fields}

The Lie algebra $\mathfrak{G}$ associated to the noise terms can be computed directly from the quantum operators. Indeed,
\begin{equation}\label{eq:Gcommute}
[G_{L_j},\, G_{L_k}] = G_{[L_j,\,L_k]}
\end{equation}
i.e.~the Lie bracket between the vector fields translates into the commutator of the associated quantum operators.
The simple proof is included in the proof of Prop.B.6. Lie brackets with drift terms are not (always) as easy to compute. Some formulas are provided later.\\

A second question is to decide, directly from the quantum master equation, when a set of vector fields is linearly dependent. By linearity of $G_L$ in $L$ and $L^\dagger$, it is clear that if $L_1,L_2,L_3$ are linearly dependent with \emph{real} coefficients, then also $G_{L_1},G_{L_2},G_{L_3}$ are linearly dependent. This condition is however not necessary. For instance, $L=I$ would give $G_I = 0$ which is collinear to any other vector field. Thus this requires a bit more analysis.

We here only consider vector fields of the form $G_L$ with $L$ independent of $\rho$. 
In general, vector fields resulting from brackets with the drift $F$ need not be of that form, i.e.~they can be equivalent to $G_L(\rho)$ with $L$ depending on $\rho$, and this may require further investigation towards linear dependencies. Note however that the results on linear independence with $L$ independent of $\rho$, of course remain valid for vector fields of the type $G_{g(\rho) L}$ with $g(\rho)$ a scalar real function, since such $\rho$-dependence does not affect the direction of the vector field. The analysis behind the present paper never had to consider other cases than this, except once for Proposition 7, where we have then analyzed the vector field directly in Bloch sphere coordinates.

It is easy to check that $G_{L+\alpha I} = G_L$ for any operator $L$ and any $\alpha \in \mathbb{C}$ (although the same is \emph{not} true for $F_L$). Therefore we assume $\text{trace}(L)=0$ in the following statements, towards better readibility. When using these results in the main paper we will thus always reduce the trace of the considered operators.

It is also important to note that we consider criteria for vector fields to be linearly dependent \emph{on a dense subset of the Bloch sphere}. E.g.~when one restricts the system to start on the surface $x^2+y^2+z^2=1$ of the Bloch sphere, the following conditions are \emph{not} necessary, and the conclusions that make use of this result must be revised.
\vspace{2mm}

\noindent \textbf{Proposition B.4:} \emph{Restricting to $\text{trace}(L_i)=0$, we have that $G_{L_1}$ and $G_{L_2}$ are linearly dependent at each $\rho$ if and only if $L_2 = \alpha L_1$ for some $\alpha \in \mathbb{R}$.}

\noindent \underline{Proof:} The `if' direction is just a check. For the `only if', we first observe that collinearity of the vector fields at $\rho = I/\text{trace}(I)$ requires $L_1+L_1^\dagger$ parallel to $L_2+L_2^\dagger$, i.e.~$L_k = \alpha_k A + i\,B_k$, $k=1,2$ with $A,B_1,B_2$ hermitian operators and $\alpha_1,\alpha_2 \in \mathbb{R}$.
\newline $\bullet$ Assume $\alpha_2 \neq 0$. Write $\rho = I + \epsilon \bar\rho$, write the collinearity condition
$G_{L_1}(\bar\rho) = g(\bar\rho)G_{L_2}(\bar\rho)$ for some scalar function $g(\bar\rho)$ (note that $G_{L_2}$ does not vanish in the neighborhood of $\rho=I/\text{trace}(I)$), and expand $g(\bar\rho)$ in powers of $\epsilon$. The first order term requires the direction of $[\alpha_2 B_1- \alpha_1 B_2,\bar\rho]$ to be independent of $\bar\rho$. For $\alpha_2 \neq 0 \neq \alpha_1$, this implies $B_1/\alpha_1 = B_2/\alpha_2$ hence in fact $L_1/\alpha_1 = L_2/\alpha_2$. For $\alpha_2 \neq \alpha_1=0$, this requires $B_1=0$ hence in fact one vector field is absent.
\newline $\bullet$ For $\alpha_1=\alpha_2=0$ we get the condition $[B_1,\rho]$ parallel to $[B_2,\rho]$ for all $rho$. There are several ways to show that also this must imply $B_1 = \alpha B_2$ for some $\alpha \in \mathbb{R}$.\hfill $\square$\\

With three vector fields, as the result appears to be nontrivial, we specialize to the qubit case, where we can write a general traceless operator as:
\begin{equation}\label{eq:zexp}
L_k = \alpha_k \sigma_x + \beta_k \sigma_y + \gamma_k \sigma_z +  i\tilde\alpha_k \sigma_x + i\tilde\beta_k \sigma_y + i\tilde\gamma_k \sigma_z \, .
\end{equation}
We then define $v_k^r := (\alpha_k,\beta_k,\gamma_k) \in \mathbb{R}^3$, $v_k^i := (\tilde\alpha_k,\tilde\beta_k,\tilde\gamma_k) \in \mathbb{R}^3$ and $v_k = (v_k^r, v_k^i) \in \mathbb{R}^6$.
We can then have linear dependencies among $G_{L_k}$ although the $L_k$ are linearly independent.
\vspace{2mm}

\noindent \textbf{Proposition B.5:} \emph{Three operators $L_1$, $L_2$, $L_3$ for a qubit system lead to vector fields $G_{L_k}$ whose directions are linearly dependent at each $\rho$, if and only if the operators' expressions $v_k$ according to \eqref{eq:zexp} satisfy one of the following conditions.
\begin{itemize}
\item Option 1: the vectors $v_1,v_2,v_3$ are linearly dependent;
\item Option 2: $v_1^r=v_2^r=v_3^r=0$, while $v_1^i,v_2^i,v_3^i$ can be arbitrary;
\item Option 3: $v_1^r,v_2^r,v_3^r$ span a two-dimensional space $\mathcal{S} \subset \mathbb{R}^3$, and there exists $\beta \in \mathbb{R}$ such that $P_{\mathcal{S}}(v_k^i) = \beta \,^\lrcorner v^r_k$ for $k=1,2,3$; here $P_{\mathcal{S}}$ denotes the orthogonal projection onto $\mathcal{S}$ and $\,^\lrcorner v$ is the vector $v$ rotated clockwise (for some convention) by $\pi/2$ in $\mathcal{S}$.
\end{itemize}
(Note that if $v_1^r,v_2^r,v_3^r$ span a one-dimensional space, then linear dependence of the $G_{L_k}$ requires linear dependence of  $v_1,v_2,v_3$.)}
\vspace{2mm}

\noindent \underline{Proof:} The most direct way to reach the conclusion is to write the vectors fields in Bloch coordinates $x,y,z$:
\begin{eqnarray}\label{eq:gx}
(G)_x & = & \alpha - \tilde\beta z + \tilde\gamma y -(\alpha x + \beta y + \gamma z)\, x\\
\nonumber (G)_y & = & \beta - \tilde\gamma x + \tilde\alpha z -(\alpha x + \beta y + \gamma z)\, y\\
\nonumber (G)_z & = & \gamma + \tilde\beta x - \tilde\alpha y -(\alpha x + \beta y + \gamma z)\, z \, .
\end{eqnarray}
From this we immediately see that at $x=y=z=0$ the vector field is $v^r$, so to have a two-dimensional vector field in the neighborhood of $x=y=z=0$ we need the $v_k^r$ to be linearly dependent.

When $v_1^r=v_2^r=v_3^r=0$, we get a purely rotating vector field, tangent to the surfaces $x^2+y^2+z^2=$constant; this is option 2. If we are not in this situation, without loss of generality, we can assume $v_1 := (1,0,0,\tilde\alpha_1,\tilde\beta_1,0)$. Since $G_{L_k}$ is linear in the coefficients of $v_k$, taking linear combinations of the $v_k$ will not change the linear dependencies in the $G_{L_{k}}$. (In particular, if $v_3$ can be expressed as a linear combination of $v_1,v_2$, then also $G_{L_3}$ can be expressed at any point as the same linear combination of $G_{L_1},G_{L_2}$, which directly gives option 1; we also recover this from the following general case.)
We can thus take such linear combinations as long as we keep track of them to re-express the original conditions on $v_1,v_2,v_3$. Given the form of $v_1$, we can make $\alpha_2 = \alpha_3 = 0$ possibly after transforming $v_2,v_3$ by linear combination with $v_1$. Then given linear dependency of the $v^r_k$ we can make $v_3^r = 0$.

From there, we obtain the result by explicitly examining linear dependence between the three instances of \eqref{eq:gx}. Write out the determinant of the 3$\times$3 matrix for the general case, this gives a polynomial in $x,y,z$ whose coefficients must all equal zero. Manageable conditions on the parameters are obtained by evaluating the polynomial at smartly selected points e.g.~$x=y=z$, and this gives the announced result.
\hfill $\square$\\

Let us briefly comment about these options.
\begin{itemize}
\item In the second option, all the $L_k$ would be skew-Hermitian, and the resulting dynamics is a pure rotation of the Bloch vector; one then easily understands that if those were the only terms in the dynamics, whatever noises are applied, the system would stay on a manifold of constant purity.
\item An example of the third case would be $L_1 = \sigma_x+ \beta i \sigma_z$, $L_2 = \sigma_z - \beta i \sigma_x$ and $L_3 = i\, \sigma_y$; this is general modulo basis change and taking real linear combinations. One checks that this set is indeed closed under commutators, i.e.~$[L_i,L_j] \in \text{span}\{ \, L_1,L_2,L_3 \, \}$ for all $i,j$. In fact, both $G_{L_3}$, and $\cos\phi \, G_{L_1}+\sin\phi \, G_{L_2}$ for a point where $x\cos\phi + z\sin\phi = 0$, imply rotations around the $y$ axis of the Bloch sphere. The second motion direction is given by
$\cos\phi \, G_{L_2} - \sin\phi \, G_{L_1}$. One then sees that $dc$ is not influenced by noise for
\begin{equation}\label{eq:StrSurf}
c = \frac{1- (x^2 + y^2 + z^2)}{(y+\beta)^2} \, .
\end{equation}
In Section \ref{sssec:1D}, example 1.b, this situation appears for an actual deterministic submanifold --- i.e.~in combination with the associated drift fields --- when $\eta=1$.
\end{itemize}

\subsubsection*{B.2 Stratonovitch vs.~It\^o}

The quantum master equation must be understood in the It\^o sense, while the correspondence between Wiener processes and controls in Prop.~A.0 holds in the Stratonovitch sense. The translation between these two formalisms involves a correction on the drift term:
\begin{eqnarray}
\nonumber \text{It\^o} & & dx_t = F(x_t) \, dt + \sum_{j=1}^m \, G_j(x)\, dW^{(j)}_t \, , \\
\nonumber &  & \;\;\;\;\; \Updownarrow\\
\nonumber\text{Strato.} & & dx_t = \left(F(x_t) + \sum_{j=1}^m D_j(x_t)\right) \, dt + \sum_{j=1}^m \, G_j(x)\, \circ\, dW^j_t \\
\label{eq:IvsS}  & & \text{with } \; D_j = - \frac{1}{2} \sum_{\ell=1}^N \, \frac{\partial G_j}{\partial x_\ell} (G_j)_\ell(x)\, ,
\end{eqnarray}
where $(G_j)_\ell$ denotes the component $\ell$ of the vector $G_j$. In general, confusing It\^o with Stratonovich would not give the same result for the dimension of the support.\footnote{For instance if $F=0$, $m=1$ it is clear that $\mathfrak{G}_F = \text{span}\{G_1\}$ i.e.~the controlled system \eqref{eq:FG} would move in one dimension, the control only determines the speed of evolution in time. The same would be true for the Stratonovitch setting \eqref{eq:FGst}, with the Wiener process just randomly varying the speed of evolution. However if this $F$ and $G_1$ were associated to an It\^o setting, then the associated Stratonovitch equation might imply motion in several dimensions. Indeed, take e.g.~$(G_1)_k(x) = (x_k)^2$, one computes that $[D_1,\,G_1]_k \propto (x_k)^4$. For almost all $x \in \mathbb{R}^N$, this vector field is not parallel to $G_1$.}

For the quantum master equation, we can give the explicit form directly for the commutation of a Stratonovitch-corrected drift with a $G$-type vector field.
\vspace{2mm}

\noindent \textbf{Proposition B.6:} \emph{In the quantum formalism, the Lie bracket between drift --- translated to Stratonovich form --- and control terms is given by:
\begin{eqnarray}
\nonumber [F_{L_j}+D_{L_j},\,G_{L_k}](\rho) & = &
 (1-\eta_j) \, \left([L_j,L_k]\rho L_j^\dagger  - \mathrm{trace}([L_j,L_k]\rho L_j^\dagger)\rho \quad + h.c. \right) \\
\nonumber  & & + (1-\eta_j)\, G_{Q_1}(\rho) \\
\label{eq:GQrho}  & & + \eta_j \, (u.i. \;\;\; + G_{Q_2})(\rho) \\
\nonumber \text{with } \;\;\; Q_1 & = & \tfrac{1}{2}\, [L_k,L_j^\dagger L_j] \;\; ; \\
\nonumber Q_2 & = & \tfrac{1}{2}\, \left[L_k\,,\, (L_j^\dagger+L_j)\, L_j \right] \;\; ;
\end{eqnarray}
$h.c.$ and $u.i.$ denoting resp.~hermitian conjugate, and unimportant terms towards Prop.A.3.}
\vspace{2mm}

\noindent \underline{Proof:}
We will assume $\eta_j\neq 0$, else the treatment basically simplifies to get the same result. From \eqref{eq:IvsS}, the Stratonovitch form of the quantum master equation is obtained by replacing $F_{L_j}$ by
$(1-\eta_j)\, F_{L_j} + \eta_j\, (F_{L_j} + \tfrac{1}{\eta_j} D_{L_j})$  with
\begin{eqnarray*}
F_{L_j} + \tfrac{1}{\eta_j} D_{L_j} & = & -\frac{1}{2} (L_j^\dagger L_j \rho + \rho L_j^\dagger L_j + (L_j)^2 \rho + \rho (L_j^\dagger)^2\, ) \\
& & + \frac{1}{2} \mathrm{trace}(L_j^\dagger L_j \rho + \rho L_j^\dagger L_j + (L_j)^2 \rho + \rho (L_j^\dagger)^2\, ) \, \rho \\
& & + \mathrm{trace}(L_j \rho + \rho L_j^\dagger)(L_j \rho + \rho L_j^\dagger) - \left(\mathrm{trace}(L_j \rho + \rho L_j^\dagger) \right)^2\, \rho \, .
\end{eqnarray*}
For a term $f(\rho)$ linear in $\rho$, we have $\Delta_{g} f \, \vert_{\rho} = f(g(\rho))$ where $\Delta_g$ denotes the Lie derivative along any vector field $g$ tangent to the state space at $\rho$. For nonlinear terms, one just applies the Leibnitz rule. E.g.~working out
\begin{eqnarray*}
\Delta_{g} G_L \, \vert_{\rho} & = & L \, g(\rho) + g(\rho) L^\dagger \\
& & - \mathrm{trace}(L g(\rho) + g(\rho) L^\dagger)\, \rho \\
& & - \mathrm{trace}(L \rho + \rho L^\dagger)\, g(\rho)
\end{eqnarray*}
leads directly to \eqref{eq:Gcommute}. Applying a similar procedure to $F_{L_j} + \tfrac{1}{\eta_j}\, D_{L_j}$ leads, after some tedious but uncomplicated computations, to
\begin{equation*}
[F_{L_j}+ \tfrac{1}{\eta_j}\, D_{L_j},\,G_{L_k}] = G_{Q(\rho)} \quad \text{with}
\end{equation*}
\begin{eqnarray}\label{eq:MoreGeneral}
Q(\rho) & = & \left[L_k\,,\, \left(\tfrac{L_j^\dagger+L_j}{2} - \mathrm{trace}(L_j \rho + \rho L_j^\dagger)\right)\, L_j \right] \\
& + & \left(\mathrm{trace}\left((L_j+L_j^\dagger)(L_k \rho + \rho L_k^\dagger)\right) - \mathrm{trace}(L_j \rho+\rho L_j^\dagger)\mathrm{trace}(L_k\rho+\rho L_k^\dagger) \right) \, L_j \, .
\end{eqnarray}
The second line contains an operator in the form $g(\rho)\, L_j$ with $g$ a real scalar function of $\rho$. Thus the associated vector field is parallel to the already present vector field $G_{L_j}$. Similarly, the first line contains a term of the form $g(\rho)[L_k,L_j]$, whose associated vector field is already in the Lie algebra $\mathfrak{G}$ involving $G_{L_j}$ and $G_{L_k}$, see the formula \eqref{eq:Gcommute}.

One easily checks that these terms do not affect the Lie algebra at higher orders. 
By thus dropping these unimportant terms in \eqref{eq:MoreGeneral}, we get the term in $Q_2$ of \eqref{eq:GQrho}.

The other terms in \eqref{eq:GQrho} are obtained directly by working out $[F_{L_j},\,G_{L_k}]$ with Lie derivatives using the Leibnitz rule. \hfill $\square$\\

The general formula \eqref{eq:GQrho} might be of limited use in practice, because it is not of the form $G_H(\rho)$ for a constant operator $H$. However, it allows us to treat several physically relevant examples for which the expressions simplify. A first case is when $\eta_j=1$, and only $G_{Q_2}(\rho)$ remains. Another practical corollary, which appears at least every time we check the bracket with $j=k$, is when $[L_j,L_k]=0$:
\vspace{2mm}

\noindent \textbf{Corollary B.7:} \emph{If $[L_j,L_k]=0$, then the Stratonovitch correction has no effect on the Lie bracket between drift associated to $L_j$ and control terms associated to $L_k$ in the quantum master SDE. Moreover, the Lie bracket reduces to:
\begin{equation}\label{eq:GQrhoA}
[F_{L_j}+D_{L_j},\,G_{L_k}] = G_{Q_2}(\rho) \;\; \text{with }Q_2 = \tfrac{1}{2}\, [L_k,L_j^\dagger] L_j\,.
\end{equation}}

\subsection*{C. General form for $L_1,L_2$ confining to a 2-dimensional submanifold}

For $\eta=1$, we have the following result. It goes beyond the scope of this work to provide interpretations for all those cases.\vspace{2mm}

\noindent \textbf{Proposition C.8} \emph{For $\eta=1$, besides the cases of heterodyne or homodyne measurements, the pairs of Lindblad operators $L_1,L_2$ which keep the system on a 2-dimensional submanifold of the Bloch sphere comprise up to basis changes:
\begin{eqnarray*}
[A] \;\;\; L_1 & = & \sigma_z + r_1\, I \\
L_2 & = & \cos\theta\,\sigma_z +\sin\theta\,\sigma_x + r_2 \, I \\
& & \text{ with any } \theta,r_1,r_2 \in \mathbb{R}\\
\, [B] \;\;\; L_1 & = & i\,\sigma_z + c_1\, I\\
L_2 & = & \sigma_x + r_2\, I\\
& &  \text{ with any } c_1\in \mathbb{C},\; r_2 \in \mathbb{R}\\
\, [C] \;\;\; L_1 & = & i\,\sigma_z + c_1\, I \\
L_2 & = & i\,(\cos\theta\,\sigma_z +\sin\theta\,\sigma_x) + c_2 \, I \\
& & \text{ with any } \theta \in \mathbb{R},\;\; c_1,c_2 \in \mathbb{C}\\
\, [D] \;\;\; L_1 & = & \cos\theta_1 \, \sigma_x + i \sin\theta_1 \, \sigma_y + r_1\, I\\
L_2 & = & \cos\theta_2 \, \sigma_x + \sin\theta_2 \, \sigma_z + r_2 \, I\\
& & \text{ with any } \theta_1,\theta_2,r_1,r_2 \in \mathbb{R}\\
\, [E] \;\;\; L_1 & = & \cos\theta \, \sigma_x + i \sin\theta \, \sigma_y + r_1\, I\\
L_2 & = & i \sigma_y + c_2\, I\\
& & \text{ with any } \theta,r_1 \in \mathbb{R},\;\; c_2 \in \mathbb{C}\\
\, [F] \;\;\; L_1 & = & \sigma_x + \beta_1\, i \sigma_y + r_1\, I \\
L_2 & = & \cos\theta\, \sigma_x + \sin\theta\, \sigma_z + \beta_2\, i \sigma_y + r_2\, I \\
& & \text{ with any } \theta,\beta_1,\beta_2,r_1,r_2 \in \mathbb{R}\\
\, [G] \;\;\; L_1 & = & \cos\theta_1 \, \sigma_x + i\,\sigma_y + \sin\theta_1\, i\, I\\
L_2 & = & \cos\theta_2\, \sigma_x + i\,(\cos\phi\, \sigma_y+\sin\phi\, \sigma_z) + \sin\theta_2\, i\, I\\
& & \text{ with any } \theta_1,\theta_2 \in \mathbb{R} \text{ and } \phi = \theta_1-\theta_2 \, .
\end{eqnarray*}
}

\noindent \underline{Sketch of the Proof:} The main text already states conditions under which a single measurement operator will not diffuse in all directions, see Prop.5. In order to stay confined to a submanifold with two measurement operators, we must already request each $L_1$ and $L_2$ individually to imply confinement, i.e.~$L_k = U_k L'_k U_k^\dagger$, with $U_k$ unitary and $L'_k$ an operator from Prop.5 with $\eta=1$, for both $k=1,2$.
We illustrate the procedure when taking $L'_1 = e^{i\phi_1} \sigma_z$, $L'_2 = e^{i\phi_2} \sigma_z$. For the other cases, the approach is essentially the same, although the algebraic computations can be tedious.

Without loss of generality, i.e.~under unitary coordinate changes, the considered case is represented by $L_1 = e^{i\phi_1} \sigma_z + c_1 I$ and $L_2 = e^{i\phi_2} (\alpha \sigma_x + \gamma \sigma_z) + c_2 I$. To avoid cases where $G_{L_1}$ and $G_{L_2}$ are already collinear --- i.e.~covered in previous results --- we impose $\alpha = 1$. \newline
- We first require $\mathfrak{G}$ to remain two-dimensional, for which we need $G_{[L_1,L_2]}$ in the span of $G_{L_1}$ and $G_{L_2}$. From $[L_1,L_2] = e^{i(\phi_1+\phi_2)} i \sigma_y$ and Prop.B.5, we identify essentially three cases where this holds: $(L_1,L_2)=(i\sigma_z,\, i(\sigma_x+\gamma \sigma_z)$ [case (a)], or $(L_1,L_2)=(\sigma_z,\, \sigma_x+\gamma \sigma_z)$ [case (b)], or $(L_1,L_2)=(i\sigma_z,\, \sigma_x)$ [case (c)]. In all these cases, arbitrary terms $c_1 I$ and $c_2 I$ may be added to the operators.\newline
- We have $[F_{L_1}+D_{L_1},\, G_{L_1}] \in \mathfrak{G}$ and $[F_{L_2}+D_{L_2},\, G_{L_2}] \in \mathfrak{G}$ because they were already collinear with $G_{L_1}$ and $G_{L_2}$ respectively. Thus there remains to check, for the various cases above, whether $[F_{L_1}+D_{L_1},\, G_{L_2}] \in \mathfrak{G}$ and $[F_{L_2}+D_{L_2},\, G_{L_1}] \in \mathfrak{G}$. Using \eqref{eq:GQrho} for $\eta=1$, this investigation boils down to checking $Q_2 \propto [L_k,\, (L_j^\dagger+L_j)L_j]$. For case (a), we have $(L_j^\dagger+L_j)L_j=(c_j+c_j^\dagger) L_j$ hence $Q_2 \propto [L_1,L_2]$. For case (b), we have $(L_j^\dagger+L_j)L_j$ composed of terms proportional to $I$, $L_j$, and $c_j L_j$. For $c_j$ complex this yields $Q_2=Q_a+Q_b$ with $Q_a \propto \mathrm{Re}(c_j)[L_1,L_2]$, and $Q_b \propto \mathrm{Im}(r_j) \sigma_y$. The latter necessarily adds a third dimension to $\mathfrak{G}_F$; thus we need $\mathrm{Im}(r_1)=\mathrm{Im}(r_2)=0$. For case (c) we have the situation of case (a) or (b) respectively, for $j=1$ or $j=2$, thus we need $\mathrm{Im}(r_2)=0$. Then in all three cases, $\mathfrak{G}_F = \mathfrak{G}$ and indeed the system would remain confined to a two-dimensional submanifold of the Bloch sphere. \hfill $\square$\\

Towards proving Prop.6, we first select the subcases from Prop.C.8 for which $L_1$ individually (and $L_2$ individually) would keep the system on a 2-dimensional manifold also when $\eta<1$, according to Prop.5. We then compute the additional term of $[F_{L_1}+D_{L_1},\, G_{L_2}]$ and write it out in Bloch coordinates. We also write out $G_{L_1}$ and $G_{L_2}$ in Bloch coordinates. After concatenating those 3 vector fields in a 3$\times$3 matrix, checking their linear dependence amounts to checking the singularity conditions of the matrix. Doing this for the general case yields the final result.

\end{document}